\documentclass[a4paper,12pt]{article}
\ifx\TeXtures\undefined
\usepackage[psamsfonts]{amsfonts}%
\usepackage{amsmath}%
\else
\usepackage{amsmath}%
\usepackage{amsfonts}%
\usepackage{times}
\fi
\usepackage{amsthm} 
\usepackage{epsfig}                
\newcommand{\DATUM}{27-June-2008}               
\newcommand{\change}
{{\marginpar{\#}}}        
\newcommand{\comma}{\: ,}              
\newcommand{\Om}{\Omega}                

\newcommand{\one}{{\bf 1}}
\newcommand{\cA}{{\mathcal{A}}}
\newcommand{\cB}{{\mathcal{B}}}

\newcommand{\cE}{{\mathcal{E}}}
\newcommand{\cF}{{\mathcal{F}}}

\newcommand{\cH}{{\mathcal{H}}}
\newcommand{\cI}{{\mathcal{I}}}

\newcommand{\cL}{{\mathcal{L}}}         
\newcommand{\cO}{{\mathcal{O}}}         

\newcommand{\cR}{{\mathcal{R}}}
\newcommand{\cS}{{\mathcal{S}}}

\newcommand{\RR}{\mathbb{R}}            
\newcommand{\ZZ}{\mathbb{Z}}            
\newcommand{\NN}{\mathbb{N}}            
\newcommand{\CC}{\mathbb{C}}            
\newcommand{\fh}{\mathfrak{h}}         
\newcommand{\vA}{{\vec{A}}}             
 
\newcommand{\vP}{{\vec{P}}}             

\newcommand{\vnabla}{{\vec{\nabla}}}
\newcommand{\veps}{{\vec{\varepsilon}}}

\newcommand{\vk}{{\vec{k}}}

\newcommand{\vp}{{\vec{p}}}

\newcommand{\vx}{{\vec{x}}}

\newcommand{\vz}{{\vec{z}}}

\newcommand{\cirS}{\mathop{\bigcirc\kern -.73em {\scriptstyle{\rm S}}}}

\newcommand{\at}{{el}}

\newcommand{\QED}{\phantom{blablabla}\hfill\qed\newline}  
\renewcommand{\thesection}
{\Roman{section}}                      
\renewcommand{\theequation}
{\thesection.\arabic{equation}}        

\newtheorem{theorem}{Theorem}[section]         
\newtheorem{lemma}[theorem]{Lemma}             
\newtheorem{corollary}[theorem]{Corollary}     
\theoremstyle{plain}
\renewcommand{\theequation}{\thesection.\arabic{equation}}
\newcommand{\resetequ}{\setcounter{equation}{0}}
\begin{document}
\bibliographystyle{plain}
\setcounter{page}{0}
\thispagestyle{empty}

\title{Renormalized Electron Mass in Nonrelativistic QED}

\author{
J\"{u}rg Fr\"{o}hlich\thanks{also at IHES, Bures-sur-Yvette} \\      
\small{Institute~of~Theoretical Physics; ETH Z\"{u}rich;} \\[-1ex] 
\small{CH-8093 Z\"{u}rich, Switzerland }\\
\small{(juerg@itp.phys.ethz.ch)} 
\and
Alessandro Pizzo\\
\small{Department of Mathematics, University of California Davis;} \\[-1ex] 
\small{One Shields Avenue, Davis, California 95616, USA }\\
\small{(pizzo@math.ucdavis.edu)}}
\date{\DATUM}

\maketitle

\begin{abstract}
Within the framework of nonrelativistic QED, we prove that, for small values of the
coupling constant,  the energy function, $E_{\vP}$,  of a dressed electron is twice
differentiable in the momentum $\vP$ in a neighborhood of $\vP=0$. Furthermore,
$\frac{\partial^2E_{\vP}}{(\partial |\vP|)^2}$ is bounded from below by a
constant larger than zero. Our results  are proven with the help of \emph{iterative analytic perturbation theory}.
\end{abstract}

\thispagestyle{empty}
\newpage
\setcounter{page}{1}
\section{Description of the problem, definition of the model,  and outline of
  the proof} 
In this paper, we study problems connected with the renormalized
electron mass in a model of quantum electrodynamics (QED) with nonrelativistic matter. We are interested in rigorously controlling  radiative
corrections to the electron mass caused by the interaction of the electron with the
\emph{soft} modes of the quantized electromagnetic field. The
model describing interactions between nonrelativistic, quantum-mechanical charged matter and the quantized radiation field
at low energies (i.e., energies smaller than the rest energy
of an electron) is the ``standard
model'', see \cite{Fierz-Pauli}. In this paper, we consider a system consisting of a
single spinless electron, described as a nonrelativistic particle that is
minimally coupled to the quantized radiation field, and photons. Electron spin
can easily be included in  our description without substantial complications.
\\
The physical system studied in this paper exhibits space translations invariance. The Hamiltonian, $H$,
generating the time evolution, commutes with the vector operator, $\vP$,
representing the total momentum of the system, which generates space
translations. If an infrared regularization, e.g., an infrared cutoff $\sigma$ on the photon
frequency, is imposed on the interaction Hamiltonian, there exist
single-electron or dressed one-electron states, as long as their momentum
is smaller than the bare electron mass, $m$, of the electron. This means that a
notion of mass shell in the energy momentum spectrum is meaningful for
velocities $|\vP|/m$ smaller than the speed of light $c$; (with $c\equiv m
\equiv 1$ in our units). Vectors $\{\Psi^{\sigma}\}$ describing dressed
one-electron states are normalizable vectors in the Hilbert space $\cH$ of
pure states of the system. They are characterized as solutions of the equation
\begin{equation}\label{eq:I.1}
H^{\sigma}\Psi^{\sigma}\,=E^{\sigma}_{\vP}\Psi^{\sigma}\,,\, |\vP|<1
\end{equation}
where $H^{\sigma}$ is the Hamiltonian with an infrared cutoff $\sigma$ in the
interaction term and $E^{\sigma}_{\vP}$, the energy of a dressed electron, is
a function of the 
momentum operator $\vP$. If in the joint spectrum of  the components of $\vP$
the support of the vector $\Psi^{\sigma}$ is contained in a ball
centered at the origin and of radius less than $1\,\equiv mc$ then
Eq. (\ref{eq:I.1}) has solutions; see
\cite{ChenFroehlichPizzo2}. Since $[H,\vP]=0$,
Eq. (\ref{eq:I.1}) can be studied for the fiber vectors,
$\Psi_{\bar{\vP}}^{\sigma}$, corresponding to a value, $\vP$, of the total
momentum; (both the total momentum operator and points in its spectrum will
henceforth be denoted by $\vP$ -- without danger of confusion). Thus we
consider the equation 

\begin{equation}\label{eq:I.2bis}
H^{\sigma}_{\vP}\Psi_{\vP}^{\sigma}\,=E^{\sigma}_{\vP}\Psi_{\vP}^{\sigma}\,,
\end{equation}
where $H^{\sigma}_{\vP}$ is the fiber Hamiltonian at fixed total momentum
$\vP$,  and
$E_{\vP}^{\sigma}$ is the value of the function $E^{\sigma}_{\vz}$ at the point $\vz\equiv \vP$.
  Physically, states $\{\Psi^{\sigma}\}$ solving Eq.~(\ref{eq:I.1}) describe
  a freely moving electron in the absence of asymptotic photons. 
\\
It is an essential aspect of the ``infrared catastrophe'' in QED that
Eq.~(\ref{eq:I.1}) does not have any normalizable solution in the limit where
the infrared cut-off $\sigma$ tends to zero, and the underlying dynamical picture of a freely moving electron
breaks down; see \cite{ChenFroehlichPizzo1}. Nevertheless,  the limiting behavior of the
function $E^{\sigma}_{\vP}$ is of great interest for the following reasons.

\noindent
As long as $\sigma>0$, a natural definition of the renormalized electron mass,
$m_r$, is given by the formula
\begin{equation}\label{eq:I.2}
m_r(\sigma):=\big[\frac{\partial^2E^{\sigma}_{|\vP|}}{(\partial |\vP|)^2} |_{\vP=0}\big]^{-1}\,.
\end{equation}
(Note that $E^{\sigma}_{\vP}\equiv E^{\sigma}_{|\vP|}$ is invariant under rotations).
Equation~(\ref{eq:I.2})  is expected to remain meaningful in the limit
$\sigma\to0$. In particular, the quantity on the R.H.S. of Eq.~(\ref{eq:I.2})
is expected to be positive and bounded from above uniformly in the infrared
cutoff $\sigma$.

\noindent
More importantly, one aims at mathematical control of the function
\begin{equation}
m_r(\sigma,| \vP|):=\big[\frac{\partial^2E^{\sigma}_{|\vP|}}{(\partial |\vP|)^2} \big]^{-1}
\end{equation}  
in a full neighborhood, $\cS$, of $\vP=0$, corresponding to a slowly moving
electron; (i.e., in the nonrelativistic regime). When combined with a number
of other spectral properties of the Hamiltonian of nonrelativistic QED the condition 
\begin{equation}\label{eq:I.5}
\frac{\partial^2E^{\sigma}_{|\vP|}}{(\partial |\vP|)^2} > 0\quad,\quad\vP\in\cS\,,
\end{equation}  
uniformly in $\sigma>0$, suffices to yield a consistent scattering  picture in
the limit when $\sigma\to0$ in which the electron exhibits
\emph{infraparticle} behavior. In fact, (\ref{eq:I.5}) is a crucial ingredient
in the analysis of Compton scattering presented in \cite{Pizzo2005}, \cite{ChenFroehlichPizzo1}.
\\

\noindent
\emph{Main results}

\noindent
Assuming the coupling constant,
$\alpha$, small enough, the following results follow.

\noindent
1) The function 
\begin{equation}\label{eq:I.6}
\Sigma_{|\vP|}:=\lim_{\sigma\to0}\frac{\partial^2E^{\sigma}_{|\vP|}}{(\partial|\vP|)^2}
\end{equation}
 is well defined for $\vP\in\cS:=\{\vP\,|\,|\vP|<\frac{1}{3}\}$; furthermore, it
 is H\"older-continuous in $\vP$.

\noindent
2) The function 
\begin{equation}
E_{\vP}:=\lim_{\sigma\to0}E_{\vP}^{\sigma}
\end{equation}
is twice differentiable in $\vP\in\cS$ and
\begin{equation}\label{eq:I.6bis}
\frac{\partial^2E_{|\vP|}}{(\partial |\vP|)^2}= \Sigma_{|\vP|}\,.
\end{equation}
 
\noindent
3)
\begin{equation}\label{eq:I.7}
\lim_{\alpha\to0}\frac{\partial^2E^{\sigma}_{|\vP|}}{(\partial|\vP|)^2 } =\frac{1}{m}\quad,\quad\vP\in\cS\,,
\end{equation}  
uniformly in $\sigma$, where  $m$ is the bare electron mass.

\noindent
(Our results can be extended to a region $\cS$ (inside the unit ball) of radius larger than $\frac{1}{3}$.)
\\
We wish to mention some related earlier results. Using \emph{operator-theoretic
renormalization group methods}, results (\ref{eq:I.6}) and (\ref{eq:I.7}) have been proven in
\cite{ChenBachFroehlichSigal} for the special value $\vP=0$. The point $\vP=0$
is exceptional, because the
Hamiltonian $H_{\vP}$ is \emph{infrared regular} at $\vP=0$; it has a
normalizable ground state. Thomas Chen (see  \cite{Chen}) has established the results in (\ref{eq:I.6bis}), (\ref{eq:I.7}) (using smooth infrared cut-offs)  by a highly non-trivial extension of the analysis of
\cite{ChenBachFroehlichSigal} to arbitrary momenta $\vP\in\cS$. 

\noindent
The procedure presented in our paper relies on \emph{iterative analytic perturbation theory} (see Section \ref{SectII} where this tool is recalled) that makes our proof substantially different and much shorter in comparison to a renormalization group approach.  The main feature is a more transparent treatment of the so called \emph{marginal} terms of the interaction,  where an essential role is played by explicit Bogoliubov transformations that transform the infrared representations of the CCR of photon creation- and annihilation operators determined by dressed one-particle states of fixed momentum $\vP(\neq 0)$ back to the Fock representation. The use of these Bogoliubov transformations is  a crucial device in our fight against the infrared problem. The way in which we are using them is new, at least in the context of mathematically rigorous results on the infrared problem in QED. 

\noindent
In our paper, the regularity properties of $E_{\vP}$ come with an explicit control of the asymptotics of the fiber  ground state eigenvectors $\Psi^{\sigma}_{\vP}$  as $\sigma$ tends to zero. (This improves earlier results in \cite{ChenFroehlichPizzo2}.)  Along the lines of \cite{BachFroehlichPizzo}, these results are  preparatory to developing an infrared finite algorithm for the asymptotic expansion of the renormalized electron mass in powers and, probably, logarithms  of  the finestructure constant $\alpha$,  up to an arbitrarily small remainder term.  However, the expansion in the coupling constant $\alpha$ is not  studied in this paper.  

\noindent
With regard to \emph{ultraviolet} corrections to the
electron mass in nonrelativistic QED models, we refer the reader to \cite{Lieb-Loss}, \cite{Lieb-Loss2},
\cite{Hiroshima-Spohn}, and \cite{Hainzl-Seiringer}.

\noindent
In Section \ref{Definition} below, the model is defined rigorously. Then, for the convenience of the reader,  in Section \ref{Outline} we outline the key ideas of the proof and present the organization of the remaining sections of the paper.
\subsection{Definition of the model} \label{Definition}

\emph{Hilbert space}

\noindent
The Hilbert space of pure state vectors of a system
consisting of one non-relativistic electron interacting with the quantized
electromagnetic field is given by
\begin{equation} \label{eq-I-1}
\cH \; := \; \cH_\at \, \otimes \, \cF  \, ,
\end{equation}
where $\cH_\at = L^2(\RR^3)$ is the Hilbert space for a
single Schr\"odinger electron; for expository convenience, we neglect
the spin of the electron.
The Hilbert space, $\cF$, used to describe the states of the transverse
modes of the quantized electromagnetic field (the \emph{photons}) in
the Coulomb gauge is given by the Fock space
\begin{equation} \label{eq-I-2}
\cF \ := \ \bigoplus_{N=0}^\infty \cF^{(N)} \comma \hspace{6mm}
\cF^{(0)} = \CC \, \Om \comma
\end{equation}
where $\Om$ is the vacuum vector (the state of the electromagnetic
field without any excited modes), and
\begin{equation} \label{eq-I-3}
\cF^{(N)} \ := \ \cS_N \, \bigotimes_{j=1}^N \fh \comma \hspace{6mm}
N \geq 1 \comma
\end{equation}
where the Hilbert space $\fh$ of state vectors of a single photon is
\begin{equation} \label{eq-I-4}
\fh \ := \ L^2( \RR^3 \times \ZZ_2 ) \,.
\end{equation}
Here, $\RR^3$ is momentum space, and $\ZZ_2$ accounts for the
two independent transverse polarizations (or helicities) of a
photon.  In (\ref{eq-I-3}), $\cS_N$ denotes the orthogonal
projection onto the subspace of $\bigotimes_{j=1}^N \fh$ of totally
symmetric $N$-photon wave functions, which accounts for the fact
that photons satisfy Bose-Einstein statistics. Thus, $\cF^{(N)}$ is
the subspace of $\cF$ of state vectors corresponding to configurations of exactly
$N$ photons.
\\

\noindent
\emph{Units}

\noindent
In this paper, we employ units such that Planck's constant $\hbar$,
the speed of light $c$, and the
mass of the electron $m$ are equal to 1.
\\

\noindent
\emph{Hamiltonian}

\noindent
The dynamics of the system is generated by the Hamiltonian
\begin{equation} \label{eq-I-6}
H \; := \; \frac{\big(-i\vnabla_{\vx} \, + \, \alpha^{1/2} \vA(\vx)
\, \big)^2}{2} \, + \, H^{f}\,.
\end{equation}
The (three-component) multiplication operator $\vx\in\RR^3$ represents the position of
the electron. The electron momentum operator is given by
$\vp=-i\vnabla_\vx$. Furthermore, 
$\alpha>0$ is the fine structure constant (which, in this paper, plays
the r\^ole of a small parameter), and $\vA(\vx)$ denotes the vector
potential of the transverse modes of the quantized electromagnetic
field in the \emph{Coulomb gauge},
\begin{equation} \label{eq-I-7}
\vnabla_\vx \cdot \vA (\vx) \ = \ 0 \, ,
\end{equation}
cutoff at high photon frequencies.

$H^f$ is the Hamiltonian of the quantized, free
electromagnetic field. It is given by
\begin{equation} \label{eq-I-10}
    H^f \; := \;  \sum_{\lambda = \pm} \int d^3k \; |\vk| \,
    a^*_{\vk,\lambda} \,  a_{\vk, \lambda} \comma
\end{equation}
where $a^*_{\vk, \lambda}$ and $a_{\vk, \lambda}$ are the usual
photon creation- and annihilation operators satisfying the
canonical commutation relations
\begin{eqnarray}
\label{eq-I-12}
    [a_{\vk, \lambda} \, , \, a^*_{\vk', \lambda'}] & = &
    \delta_{\lambda \lambda'} \, \delta (\vk - \vk') \comma
    \\
    \label{eq-I-11}
    [a^\#_{\vk, \lambda} \, , \, a^\#_{\vk', \lambda'}] & = & 0
\end{eqnarray}
for $\vk, \vk' \in \RR^3$ and $\lambda,\lambda' \in \ZZ_2\equiv\{\pm\}$, where $a^\#=a$ or $a^*$.
The vacuum vector $\Omega\in\cF$ is characterized by the condition
\begin{equation}
    a_{\vk, \lambda} \, \Om \; = \; 0 \comma  \label{eq-I-13}
\end{equation}
for all $\vk\in \RR^3$ and $\lambda \in \ZZ_2\equiv\{\pm\}$.

The quantized electromagnetic vector potential is given by
\begin{eqnarray} \label{eq-I-14}
    \vA(\vx) \; := \;
    \sum_{\lambda = \pm} \int_{\cB_{\Lambda}} \frac{d^3k}{\sqrt{ |\vk| \,}}\,
    \big\{ \veps_{\vk, \lambda} e^{-i\vk \cdot \vx}
    a^*_{\vk, \lambda} \, + \, \veps_{\vk, \lambda}^{\,\,*}
    e^{i\vk \cdot \vx} a_{\vk, \lambda} \big\} \comma
\end{eqnarray}
where $\veps_{\vk, -}$, $\veps_{\vk, +}$ are photon polarization
vectors, i.e., two unit vectors in $\RR^3 \otimes\CC$ satisfying
\begin{equation} \label{eq-I-15}
    \veps_{\vk, \lambda}^{\,*} \cdot \veps_{\vk, \mu} \; = \;
    \delta_{\lambda \mu} \comma \hspace{8mm} \vk \cdot \veps_{\vk,
    \lambda} \; = \; 0 \comma
\end{equation}
for $\lambda, \mu = \pm$. The equation $\vk \cdot \veps_{\vk,
\lambda} = 0$ expresses the Coulomb gauge condition. Moreover,
$\cB_{\Lambda}$ is a ball of radius $\Lambda$ centered at the origin
in momentum space. Here, $\Lambda$ represents an \emph{ultraviolet cutoff}
that will be kept fixed throughout our analysis. The vector
potential defined in (\ref{eq-I-14}) is thus cut off in the
ultraviolet.

Throughout this paper, it will be assumed that $\Lambda\approx 1$
(the rest energy of an electron), and that $\alpha>0$ is sufficiently small.
Under these assumptions, the Hamiltonian $H$ is selfadjoint on $D(H^0)$, i.e.,
on the domain of definition of the operator
\begin{equation}
    H^0 \; := \; \frac{(-i\vnabla_{\vx})^2}{2} \, + \, H^f \;.
\end{equation}
The perturbation $H-H^0$ is small in the sense of Kato.

The operator representing the total momentum of the system consisting of the
electron and the electromagnetic radiation field is given by
\begin{equation}
    \vP\,:=\,\vp+\vP^f \, ,
\end{equation}
with $\vp=-i\vnabla_{\vx}$, and where
\begin{equation}
    \vP^f\,:=\, \sum_{\lambda = \pm} \int  d^3k \; \vk \,
    a^*_{\vk, \lambda} \, a_{\vk, \lambda}
\end{equation}
is the momentum operator associated with the photon field.

The operators $H$ and $\vP$ are essentially selfadjoint on a common
domain, and since the dynamics is invariant under
translations, they commute, $[H,\vP]=\vec 0$.
The Hilbert space $\cH$ can be decomposed into a direct integral over the joint
spectrum, $\RR^3$,  of the three components of the momentum operator $\vP$.
Their spectral measure is
absolutely continuous with respect to Lebesgue measure, and hence we have that
\begin{equation}
\cH\,:=\,\int^{\oplus}\cH_{\vP} \, d^3P \, ,
\end{equation}
where each fiber space $\cH_{\vP}$ is a copy of Fock space $\cF$.
\\

\noindent {\bf{Remark}} \emph{Throughout
this paper, the symbol $\vP$ stands for both a vector in $\RR^3$
and the vector operator on $\cH$, representing the total momentum, depending on context.
Similarly, a double meaning is
given to arbitrary functions, $f(\vP)$, of the total
momentum operator.}
\\

We recall that vectors $\Psi\in\cH$ are given by sequences
\begin{equation}
\{\Psi^{(m)}(\vx; \vk_1,\lambda_1;\dots; \vk_m,\lambda_m)\}_{m=0}^{\infty}\,,
\end{equation}
of functions, $\Psi^{(m)}$, where $\Psi^{(0)}(\vx)\in L^2(\RR^3)$, of the electron position $\vx$ and of $m$ photon momenta
$\vk_1,\dots,\vk_m $ and helicities $\lambda_1, \dots, \lambda_m$, with the following properties:
\begin{itemize}
\item[(i)]
$\Psi^{(m)}(\vx; \vk_1,\lambda_1;\dots; \vk_m,\lambda_m)$
is totally symmetric in its $m$ arguments $(\vk_j,\lambda_j)_{j=1,\dots,m}$.
\item[(ii)]
$\Psi^{(m)}$ is square-integrable, for all $m$.
\item[(iii)]
If $\Psi$ and $\Phi$ are two vectors in $\cH$ then 
\begin{eqnarray}
\lefteqn{(\Psi\,,\,\Phi) }\\
&=&\sum_{m=0}^{\infty}\big(\sum_{\lambda_j=\pm}\,\int\,d^3x\,\prod_{j=1}^{m}\,d^3k_j\,\overline{\Psi^{(m)}(\vx;\vk_1,\lambda_1;\dots; \vk_m,\lambda_m)}\,\Phi^{(m)}(\vx;\vk_1,\lambda_1;\dots; \vk_m,\lambda_m)\big)\,.\nonumber
\end{eqnarray}
\end{itemize}
We identify a square integrable function $g(\vx)$ with the sequence 
\begin{equation}
\{\Psi^{(m)}(\vx; \vk_1,\lambda_1;\dots; \vk_m,\lambda_m)\}_{m=0}^{\infty}\,,
\end{equation}
where $\Psi^{(0)}(\vx)\equiv g(\vx)$, and $\Psi^{(m)}(\vx; \vk_1,\lambda_1;\dots; \vk_m,\lambda_m)\equiv 0$ for all $m>0$;  analogously, a square integrable function $g^{(m)}(\vx; \vk_1,\lambda_1;\dots; \vk_m,\lambda_m)$, $m\geq 1$,  is identified with the sequence
\begin{equation}
\{\Psi^{(m')}(\vx; \vk_1,\lambda_1;\dots; \vk_{m'},\lambda_{m'})\}_{m'=0}^{\infty}\,,
\end{equation}
where $\Psi^{(m)}(\vx; \vk_1,\lambda_1;\dots; \vk_m,\lambda_m)\equiv g^{(m)} $, and  $\Psi^{(m')}(\vx; \vk_1,\lambda_1;\dots; \vk_{m'},\lambda_{m'})\equiv 0$ for all $m'\neq m$. From now on, a sequence describing a quantum state with a fixed number of photons is identified with  its nonzero component wave function; vice versa, a wave function corresponds to a sequence according to the previous identification.
The elements of the fiber space $\cH_{\vP^*}$ are obtained by linear combinations of the (improper) eigenvectors  of the total momentum operator $\vP$ with eigenvalue $\vP^*$, e.g., the plane wave $e^{i\vP^*\cdot\vx}$ is the eigenvector describing a state with an electron and no photon. 

\noindent
Given any $\vP\in\RR^3$, there is an isomorphism, $I_{\vP}$,
\begin{equation}
    I_{\vP}\,:\,\cH_{\vP}\,\longrightarrow\,\cF^{b}\,,
\end{equation}
from the fiber space $\cH_{\vP}$ to the Fock space $\cF^{b}$, acted upon by the annihilation- and
creation operators $b_{\vk,\lambda}$, $b^*_{\vk,\lambda}$,
where
$b_{\vk,\lambda}$ corresponds to $e^{i\vk\cdot\vx}  a_{\vk,\lambda}$, and
$b_{\vk,\lambda}^*$ to $e^{-i\vk\cdot\vx} a_{\vk,\lambda}^* $, and
with vacuum $\Omega_{f}:=I_{\vP}(e^{i\vP\cdot\vx})$.
To define $I_{\vP}$ more precisely, we consider a vector
$\psi_{(f^{(n)};\vP)}\in \cH_{\vP}$
with a definite total momentum describing an electron and $n$ photons. Its wave function in the variables
$(\vx;\vk_1,\lambda_1;\dots,\vk_n,\lambda_n)$ is given by
\begin{equation}
 e^{i(\vP-\vk_1-\cdots-\vk_n)\cdot\vx}f^{(n)}(\vk_1,\lambda_1;\dots;\vk_n,\lambda_n)
\end{equation}
where $f^{(n)}$ is totally symmetric in its $n$ arguments.
The isomorphism $I_{\vP}$ acts by way of
\begin{eqnarray}
    \lefteqn{I_{\vP}\big( e^{i(\vP-\vk_1-\cdots-\vk_n)\cdot\vx}f^{(n)}(\vk_1,\lambda_1;\dots;\vk_n,\lambda_n)\big)}
    \\
    &= &\frac{1}{\sqrt{n!}}\sum_{\lambda_1,\dots,\lambda_n}\int \, d^3k_1\dots d^3k_n \,f^{(n)}(\vk_1,\lambda_1;\dots;\vk_n,\lambda_n)\,
    b_{\vk_1,\lambda_1}^* \cdots
    b_{\vk_n,\lambda_n}^*  \, \Omega_f \,.\nonumber
\end{eqnarray}
Because the Hamiltonian $H$ commutes with the total momentum, it preserves the
fibers $\cH_{\vP}$ for all $\vP\in\RR^3$, i.e., 
it can be written as
\begin{equation}
    H\,=\,\int^{\oplus} H_{\vP}\,d^3P\,,
\end{equation}
where
\begin{equation}
    H_{\vP}\,:\,\cH_{\vP}\longrightarrow\cH_{\vP}\,.
\end{equation}
Written in terms of the operators $b_{\vk,\lambda}$, $b^*_{\vk,\lambda}$, and of the
variable $\vP$, the fiber Hamiltonian $H_{\vP}$ is given by
\begin{equation}\label{eq-fibHam}
    H_{\vP} \; := \; \frac{\big(\vP-\vP^f +\alpha^{1/2} \vA \big)^2}{2} \;
    + \;H^{f}\,,
\end{equation}
where
\begin{eqnarray}
    \vP^f &= & \sum_{\lambda}\, \int d^3k\, \vk\, b^*_{\vk,\lambda} \, b_{\vk,\lambda} \, ,
    \\
    H^f & =& \sum_{\lambda}\, \int d^3k \, |\vk| b^*_{\vk,\lambda}  b_{\vk,\lambda} \,,
\end{eqnarray}
and
\begin{equation}
    \vA \, := \,\sum_{\lambda}\, \int_{\cB_{\Lambda}}\,
    \frac{d^3k}{\sqrt{ |\vk| \,}} \, \big\{  \veps_{\vk,\lambda}b^*_{\vk,\lambda} 
    \, + \, \veps^{\,\,*}_{\vk,\lambda} b_{\vk,\lambda} \big\} \,.
\end{equation}
Let
\begin{equation}
    \mathcal{S}:=\lbrace\,
    \vP\in\RR^3\,:\,|\vP|<\frac{1}{3}\,\rbrace\,.
\end{equation}
In order to give a mathematically precise meaning to the constructions
presented in the following, we introduce an infrared cut-off at a photon
frequency $\sigma>0$ in the vector potential. The calculation of the second
derivative of the energy of a dressed electron -- in the following called the
``ground state energy'' -- as a function of $\vP$ in the limit where $\sigma\to0$, and for $\vP\in\cS$, represents the main problem solved in this paper.
Hence  we will, in the sequel, study the regularized fiber Hamiltonian
\begin{equation}\label{eq:H-fiber}
    H_{\vec{P}}^{\sigma} \, := \, \frac{\big(\vec{P}-\vec{P}^{f}
    +\alpha^{1/2} \vec{A}^{\sigma} \big)^2}{2} \,  + \, H^{f}\,,
\end{equation}
acting on the
fiber space $\mathcal{H}_{\vec{P}}$, for $\vec{P}\in \mathcal{S}$,
where
\begin{eqnarray}
    \vec{A}^{\sigma} \, := 
    \,\sum_{\lambda}\, \int_{\mathcal{B}_{\Lambda}\setminus \mathcal{B}_{\sigma}}\,
    \frac{d^3k}{\sqrt{ |\vk| \,}} \, \big\{   \veps_{\vk,\lambda}b^*_{\vk,\lambda}\, + \,
    \veps^{\,\,*}_{\vk,\lambda} b_{\vk,\lambda} \big\}
\end{eqnarray}
and where $\mathcal{B}_{\sigma}$ is a ball of radius $\sigma$ centered at the
origin. In the following, we will consider a sequence of infrared cutoffs  
\begin{equation}
\sigma_j \, := \, \Lambda\epsilon^{j}
\end{equation}
with $0<\epsilon<\frac{1}{2}$ and $j\in\mathbb{N}_{0}:=\NN \cup \{0\}$.
\\

\noindent
\emph{Notations}

\noindent
1) We use the notation $\| A \|_{\cH}=\|A|_{\cH}\|$ for the norm of a bounded operator $A$ acting on a Hilbert space $\cH$. Typically, $\cH$ will be some subspace of $\cF^b$.

\noindent
2) Throughout the paper, we follow conventions such that
$$  \frac{1}{2\pi i}\oint_{\gamma}\frac{1}{z}dz=-1\quad,\quad  \oint_{\gamma}\frac{1}{\bar{z}}d\bar{z}=\overline{( \oint_{\gamma}\frac{1}{z}dz)}\,,$$
where $\gamma$ is an integration path in the complex space enclosing the origin.
\subsection{Outline of the proof}\label{Outline}

Next, we outline the key ideas used in the proofs of our main results in Eqs. (\ref{eq:I.6}), (\ref{eq:I.6bis}), and (\ref{eq:I.7}).

\noindent
For $\vP\in\cS$, $\alpha$ small enough, and $\sigma>0$, $E^{\sigma}_{\vP}$ is an isolated eigenvalue of
$H_{\vP}^{\sigma}|_{\cF_{\sigma}}$; see Section II and Eq. (\ref{eq:II.4}). Because of the analyticity
of $H_{\vP}^{\sigma}$ in the variable $\vP$, it follows that
\begin{eqnarray}\label{eq:I.36}
& &\frac{\partial^2E^{\sigma}_{|\vP|}}{(\partial|\vP|)^2}=\partial_i^2E^{\sigma}_{|\vP|}|_{\vP=P^{i}\hat{i}}\\
&=&1-2 \langle \frac{1}{2\pi i}\oint_{\gamma_{\sigma}}\frac{1}{H^{\sigma}_{\vP}-z}[\partial_iH^{\sigma}_{\vP}]\frac{1}{H^{\sigma}_{\vP}-z}dz\,\Psi_{\vP}^{\sigma},\,[\partial_iH^{\sigma}_{\vP}]\Psi_{\vP}^{\sigma}\rangle|_{\vP=P^{i}\hat{i}} \,, \nonumber
\end{eqnarray}
where $\partial_i=\partial/\partial P^{i}$, $\hat{i}$ is the unit vector in
the direction $i$, $\Psi_{\vP}^{\sigma}$ is the
normalized ground state eigenvector of
$H_{\vP}^{\sigma}$ constructed in \cite{ChenFroehlichPizzo2}; $\gamma_{\sigma}$ is
a contour path in the complex energy plane enclosing $E^{\sigma}_{\vP}$ and
no other point of the spectrum of $H_{\vP}^{\sigma}$, and such that the distance of $\gamma_{\sigma}$ from
$\text{spec}\,(H_{\vP}^{\sigma})$ is of order $\sigma$. 

\noindent
At first glance, the expression on the R.H. S. of (\ref{eq:I.36})  might become  singular
 as $\sigma\to0$, because the spectral gap above $E^{\sigma}_{\vP}=\inf \text{spec}
\, (H^{\sigma}_{\vP}|_{\cF_{\sigma}})$ is of order $\sigma$. To prove that the limit
 $\sigma\to0$ is, in fact, well defined, we make use of a $\sigma$-dependent Bogoliubov
 transformation, $W_{\sigma}(\vnabla
E_{\vP}^{\sigma})$;  (see Section II, Eq. (\ref{eq:II.3})). This transformation has already been employed in
 \cite{ChenFroehlichPizzo2} to analyze mass shell properties. In fact, conjugation of $H^{\sigma}_{\vP}$ by $W_{\sigma}(\vnabla
E_{\vP}^{\sigma})$ yields an infrared regularized Hamiltonian
\begin{equation}
K_{\vP}^{\sigma}\,:=\,W_{\sigma}(\vnabla
E_{\vP}^{\sigma})H_{\vP}^{\sigma}W_{\sigma}^*(\vnabla
E_{\vP}^{\sigma})\,
\end{equation}
with the property that the corresponding ground state,
 $\Phi^{\sigma}_{\vP}$, 
has a non-zero limit, as $\sigma\to0$. The Hamiltonian $K_{\vP}^{\sigma}$ has a
``canonical form'' derived in \cite{ChenFroehlichPizzo2} (see also  \cite{Pizzo}, where a similar operator has been used in the analysis of the Nelson model):
\begin{equation}
K_{\vP}^{\sigma}\,=\,\frac{(\vec{\Gamma}^{\sigma}_{\vP})^2}{2}+\sum_{\lambda}\int_{\RR^3}|\vk|\delta_{\vP}^{\sigma}(\hat{k})b^*_{\vk,\lambda}b_{\vk,\lambda}d^3k+\cE_{\vP}^{\sigma}\,,
\end{equation}
where $\delta_{\vP}^{\sigma}(\hat{k})$ is defined in Eq. (\ref{eq-III.18}), $\cE_{\vP}^{\sigma}$ is a c-number defined in Eq. (\ref{eq-II.42}), and $\vec{\Gamma}^{\sigma}_{\vP}$
is a vector operator defined in Eq. (\ref{eq:II.40}) starting from Eqs. (\ref{eq:III.16}), (\ref{eq-III.17}), (\ref{eq:II.34bis}). By construction,
\begin{equation}\label{eq:I.39}
\langle \Phi^{\sigma}_{\vP}\,,\,\vec{\Gamma}^{\sigma}_{\vP}\,\Phi^{\sigma}_{\vP}\rangle=0\,.
\end{equation}
This is a crucial property in the proof of existence of a limit of $\Phi^{\sigma}_{\vP}$ as $\sigma\to0$.
 \\
 Eq. (\ref{eq:I.39}) is also an important ingredient in the proof of (\ref{eq:I.6}), because, by applying  the unitary operator $W_{\sigma}(\vnabla
E_{\vP}^{\sigma})$  to each term of the scalar product on the R.H.S. of (\ref{eq:I.36})
and using (\ref{eq:I.39}) (see Section \ref{SectIII}), one finds that
\begin{eqnarray}
(\ref{eq:I.36})&=&1-2\frac{1}{\|\Phi_{\vP}^{\sigma_j}\|^2}\langle \frac{1}{2\pi i}\oint_{\gamma_j}\frac{1}{K^{\sigma_j}_{\vP}-z_j}[\partial_iE_{\vP}^{\sigma_j}-(\Gamma^{\sigma_j}_{\vP})^{i}]\frac{1}{K^{\sigma_j}_{\vP}-z_j}dz_j\,\Phi_{\vP}^{\sigma_j}\,,\nonumber\\
& &\quad\quad\quad\quad\quad\quad\quad\quad\quad\quad\,,\,[\partial_iE_{\vP}^{\sigma_j}-(\Gamma^{\sigma_j}_{\vP})^{i}]\Phi_{\vP}^{\sigma_j}\rangle|_{\vP=P^{i}\hat{i}} \label{eq:I.39bis} \\
&=&1-2 \langle \frac{1}{2\pi i}\oint_{\gamma_{\sigma}}\frac{1}{K^{\sigma}_{\vP}-z}\,(\Gamma^{\sigma}_{\vP})^i\, \frac{1}{K^{\sigma}_{\vP}-z}dz\,\frac{\Phi_{\vP}^{\sigma}}{\|\Phi_{\vP}^{\sigma}\|}\,,\,(\Gamma^{\sigma}_{\vP})^i\,\frac{\Phi_{\vP}^{\sigma}}{\|\Phi_{\vP}^{\sigma}\|}\rangle|_{\vP=P^{i}\hat{i}}\,. \quad\quad\quad
\label{eq:I.40}
\end{eqnarray}
Notice that, if one starts from the two expressions on the R.H.S. of Eq.  (\ref{eq:I.36}) and on  the R.H.S. of Eq. (\ref{eq:I.39bis}) respectively, both formally expanded in powers of $\alpha^{1/2}$,  the Bogoliubov transformation can be seen as a tool to re-collect an infinite number of terms and to show a nontrivial identity, thanks only to  Eq. (\ref{eq:I.39})  and to a vanishing contour integration (see Eq. (\ref{eq:III.48})). Next, still using  Eq. (\ref{eq:I.39}), one can show that (\ref{eq:I.40})
remains uniformly bounded in $\sigma$.
\\
To see this we use the inequality
\begin{equation}\label{eq:I.41}
\big|\big\langle (\Gamma_{\vP}^{\sigma})^{i}\,\Phi_{\vP}^{\sigma}\,,\,\big(\frac{1}{K_{\vP}^{\sigma}-z}\big)^2(\Gamma_{\vP}^{\sigma})^{i}\,\Phi_{\vP}^{\sigma}\big \rangle \big|\leq \cO(\frac{1}{\alpha^{\frac{1}{2}}\, \sigma^{2\delta}})\,,
\end{equation}
for an arbitrarily small $\delta>0$, with $z\in\gamma_{\sigma}$ and $\alpha$ small enough depending on $\delta$. 
This inequality will be proven inductively (see Theorem \ref{Th.induction}) by introducing sequences of infrared cut-offs $\sigma_j$, where $\sigma_j\to0$ as $j\to \infty$. The proof by induction is combined with an improved (as compared to the result in \cite{ChenFroehlichPizzo2}) estimate of the rate of convergence of $\{\Phi^{\sigma}_{\vP}\}$ as $ \sigma\to0$ . 

  \noindent
  By telescoping, one can plug these improved estimates into (\ref{eq:I.40}) to end up with the desired uniform bound. The control of the rate of convergence  of the R.H.S. in (\ref{eq:I.40}), as $\sigma\to0$, combined with the smoothness in $\vP$, for arbitrary infrared cutoff $\sigma>0$, finally entails the H\"older-continuity in $\vP$ of the limiting quantity 
 \begin{equation}
 \Sigma_{|\vP|}\,:=\,1-\lim_{\sigma\to0} 2 \langle \frac{1}{2\pi i}\oint_{\gamma_{\sigma}}\frac{1}{K^{\sigma}_{\vP}-z}\,(\Gamma^{\sigma}_{\vP})^i\,\frac{1}{K^{\sigma}_{\vP}-z}dz\,\frac{\Phi_{\vP}^{\sigma}}{\|\Phi_{\vP}^{\sigma}\|}\,,\,(\Gamma^{\sigma}_{\vP})^i\,\frac{\Phi_{\vP}^{\sigma}}{\|\Phi_{\vP}^{\sigma}\|}\rangle|_{\vP=P^{i}\hat{i}} \,.
\end{equation}
The H\"older-continuity in $\vP$ of $ \Sigma_{|\vP|}$ and of $\lim_{\sigma\to0}\frac{\partial E^{\sigma}_{|\vP|}}{\partial |\vP|}$, combined with the fundamental theorem of calculus, imply that $E_{\vP}$ is twice differentiable and 
$\frac{\partial^2 E_{|\vP|}}{(\partial |\vP|)^2}\equiv  \Sigma_{|\vP|}$.
\\

Our paper is organized as follows.

\noindent
In Section \ref{SectII}, we recall how to construct the ground states of the Hamiltonians $H_{\vP}^{\sigma}$ and $K_{\vP}^{\sigma}$ by \emph{iterative analytic perturbation theory}. This section contains an explicit derivation of the formula of the transformed Hamiltonians and of related algebraic identities that will be used later on.

\noindent
In  Section \ref{SectIII}, we first derive inequality (\ref{eq:I.41}) and the improved convergence rate  of $\{\Phi^{\sigma}_{\vP}\}$ as $\sigma\to0$, by using some key ingredients described in Section  \ref{SectII}. Section  \ref{SectIII.1}  is devoted to an analysis of (\ref{eq:I.36}) that culminates in the following main results.
  
\noindent
\emph{\bf{Theorem}}

\noindent
\emph{For $\alpha$ small enough, $\frac{\partial^2E^{\sigma}_{|\vP|}}{(\partial|\vP|)^2}$ converges as $\sigma\to0$. The limiting function  $ \Sigma_{|\vP|}:=\lim_{\sigma\to0}\,\frac{\partial E^{\sigma}_{|\vP|}}{(\partial|\vP|)^2}$  is H\"older continuous in $\vP\in\cS$. The  limit
\begin{equation} \label{eq:IV.71}
\lim_{\alpha\to0}\Sigma_{|\vP|}=1\,
\end{equation}
holds true uniformly in $\vP\in\cS$.}

\noindent
\emph{\bf{Corollary}}

\noindent
\emph{For $\alpha$ small enough, the function $E_{\vP}:=\lim_{\sigma\to0}E_{\vP}^{\sigma}$,  $\vP\in\cS$, is twice
differentiable, and 
\begin{equation}
\frac{\partial^2E_{|\vP|}}{(\partial |\vP|)^2}= \Sigma_{|\vP|}\quad\text{for all}\quad\vP\in\cS\,.
\end{equation}}
\\

\noindent

\noindent
{\bf{Remark}}

\noindent
For the complete proof of the construction of the ground states of the Hamiltonians $H_{\vP}^{\sigma}$ and $K_{\vP}^{\sigma}$ by \emph{iterative analytic perturbation theory}, the reader is advised to consult ref. \cite{ChenFroehlichPizzo2}. 


\section{Sequences  of ground state vectors}\label{SectII}
\resetequ

In this section, we report on results contained in
\cite{ChenFroehlichPizzo2} concerning the ground
states of the Hamiltonians $H_{\vP}^{\sigma_j}$, where $\vP\in\cS$ and $j\in \NN_0$, and  the existence
of a limiting vector for the sequence of ground state vectors of the transformed Hamiltonians, $K_{\vP}^{\sigma_j}$, where the Bogoliubov transformation used to obtain $K_{\vP}^{\sigma_j}$ from $H_{\vP}^{\sigma_j}$ (derived in
\cite{ChFr}) is determined by
\begin{eqnarray}
b^{\,*}_{\vk,\lambda}&\rightarrow&W_{\sigma_j}(\vnabla E_{\vP}^{\sigma_j})b^{\,*}_{\vk,\lambda}W_{\sigma_j}^*(\vnabla E_{\vP}^{\sigma_j})\,=\,b^{\,*}_{\vk,\lambda}-\alpha^{\frac{1}{2}}\frac{\vnabla
  E_{\vP}^{\sigma_j}\cdot\vec{\epsilon}^{\,*}_{\vk,\lambda}}{|\vk|^{\frac{3}{2}}\delta_{\vP}^{\sigma_j}(\hat{k})}\\
b_{\vk,\lambda}&\rightarrow&W_{\sigma_j}(\vnabla E_{\vP}^{\sigma_j})b_{\vk,\lambda}W_{\sigma_j}^*(\vnabla E_{\vP}^{\sigma_j})\,=\,b_{\vk,\lambda}-\alpha^{\frac{1}{2}}\frac{\vnabla
  E_{\vP}^{\sigma_j}\cdot\vec{\epsilon}_{\vk,\lambda}}{|\vk|^{\frac{3}{2}}\delta_{\vP}^{\sigma_j}(\hat{k})}\,,
\end{eqnarray}
for $\vk\in \cB_{\Lambda}\setminus\cB_{\sigma_j}$, with
\begin{equation}\label{eq:II.3}
W_{\sigma_j}(\vnabla E_{\vP}^{\sigma_j})\,:=\,\exp\big(\alpha^{\frac{1}{2}}\sum_{\lambda}\int_{\cB_{\Lambda}\setminus\cB_{\sigma_j}}d^3k\frac{\vnabla
  E_{\vP}^{\sigma_j}}{|\vk|^{\frac{3}{2}}\delta_{\vP}^{\sigma_j}(\hat{k})}\cdot(\vec{\epsilon}_{\vk,\lambda}b^*_{\vk,\lambda}-h.c.)\big)\,,
\end{equation}
where $\delta_{\vP}^{\sigma_j}(\hat{k})$ is defined in (\ref{eq-III.18}).

\subsection{Ground states of the Hamiltonians $ H_{\vec{P}}^{\sigma_j}$.} \label{SectII.0}

\noindent
In  \cite{ChenFroehlichPizzo2}, the first step consists in constructing the ground states of the
regularized fiber Hamiltonians $ H_{\vec{P}}^{\sigma_j}$. As shown in \cite{ChenFroehlichPizzo2},
$   H_{\vec{P}}^{\sigma_j}$ has a unique ground state, $\Psi_{\vP}^{\sigma_j}$,
that can be constructed by \emph{iterative analytic perturbation
  theory},  as developed in \cite{Pizzo}. To recall how this method works some preliminary definitions and
results are needed:
\begin{itemize}
\item
We introduce the Fock spaces
\begin{equation}\label{eq:II.4}
\cF_{\sigma_j}:=\cF^b(L^2((\RR^3\setminus\cB_{\sigma_j})\times \ZZ_2))\,\quad,\quad\,\cF^{\sigma_j}_{\sigma_{j+1}}:=\cF^b(L^2((\cB_{\sigma_j}\setminus\cB_{\sigma_{j+1}})\times \ZZ_2))\,.
\end{equation}
Note that
\begin{equation}
\cF_{\sigma_{j+1}}\,=\,\cF_{\sigma_j}\otimes\cF^{\sigma_j}_{\sigma_{j+1}}\,.
\end{equation}
If not specified otherwise, $\Omega_f$ denotes the vacuum state in anyone of
these Fock spaces. Any vector $\phi$ in $\cF_{\sigma_j}$ can be identified with the
corresponding vector, $\phi\otimes\Omega_f$, in $\cF$, where $\Omega_f$ is the
vacuum in $\cF^{\sigma_j}_{0}$.

\item Momentum-slice interaction Hamiltonians
are defined by
\begin{equation}
\Delta H_{\vP}|^{\sigma_j}_{\sigma_{j+1}}:=\alpha^{\frac{1}{2}}\,\vnabla_{\vP}H_{\vP}^{\sigma_j}\cdot\vec{A}|^{\sigma_j}_{\sigma_{j+1}}+\frac{\alpha}{2}\,({\vec{A}}|^{\sigma_j}_{\sigma_{j+1}})^2\,,
\end{equation}
where
\begin{eqnarray}
    \vec{A}|^{\sigma_j}_{\sigma_{j+1}} \, := 
    \,\sum_{\lambda}\, \int_{\mathcal{B}_{\sigma_j}\setminus \mathcal{B}_{\sigma_{j+1}}}\,
    \frac{d^3k}{\sqrt{ |\vk| \,}} \, \big\{   \veps_{\vk,\lambda}b^*_{\vk,\lambda}\, + \,
    \veps^{\,\,*}_{\vk,\lambda} b_{\vk,\lambda} \big\}\,;
\end{eqnarray}
\item
Four real parameters, $\epsilon,\,\rho^+,\,\rho^-$, and $\mu$, will appear in our analysis. They have the properties
\begin{eqnarray}
& &0<\rho^-<\mu<\rho^+<1-C_{\alpha}<\frac{2}{3}\label{eq:III.8bis}\\
& &0<\epsilon<\frac{\rho^-}{\rho^+}\label{eq:III.9}
\end{eqnarray}
where $C_{\alpha}$, with $\frac{1}{3}<C_{\alpha}<1$, for $\alpha$ small enough, is a constant such that the inequality
\begin{equation}\label{eq:II.10}
E_{\vP-\vk}^{\sigma}>E_{\vP}^{\sigma}-C_{\alpha}|\vk|\,
\end{equation}
holds for all $\vP\in\cS$ and any $\vk\neq0$. Here $E_{\vP-\vk}^{\sigma}:=\text{inf spec}H_{\vP-\vk}^{\sigma}$. We note that $C_{\alpha}\to\frac{1}{3}$, as $\alpha\to 0$; (see Statement
($\mathcal{I}4$) of  Theorem 3.1.  in \cite{ChenFroehlichPizzo2}).
\end{itemize}
By iterative analytic perturbation theory (see \cite{ChenFroehlichPizzo2}),  one derives
the following results, valid for sufficiently small $\alpha$, depending on our choice of $\Lambda, \epsilon,\rho^-, \mu,$ and $ \rho^+$ (see also Figure 1 below):
 \begin{itemize}
\item [ ($\cA1$)]
 $E_{\vP}^{\sigma_j}$ is an isolated simple eigenvalue of $H_{\vP}^{\sigma_j}|_{\cF_{\sigma_j}}$ with spectral gap larger or equal to $\rho^-\sigma_j$. 
Furthermore,  $E_{\vP}^{\sigma_j}$ is also the ground state energy of $H_{\vP}^{\sigma_j}|_{\cF_{\sigma_{j+1}}}$, and it is an isolated simple eigenvalue of $H_{\vP}^{\sigma_j}|_{\cF_{\sigma_{j+1}}}$ with spectral gap larger or equal to $\rho^+\sigma_{j+1}$.
 \item [($\cA 2$)]
The ground-state energies $E_{\vP}^{\sigma_j}$ and $E_{\vP}^{\sigma_{j+1}}$ of the Hamiltonians $H_{\vP}^{\sigma_j}$ and $H_{\vP}^{\sigma_{j+1}}$, respectively,  (acting on the same space $\cF_{\sigma_{j+1}}$) satisfy the inequalities
\begin{equation}
0\leq E_{\vP}^{\sigma_{j+1}} \leq E_{\vP}^{\sigma_{j}}+c\,\alpha\,\sigma_j^2\,,
\end{equation}
where $c$ is independent of  $j$ and of  $\alpha$.
\item [($\cA 3$)]
The ground state vectors,  $\Psi_{\vP}^{\sigma_{j+1}}$,  of $H_{\vP}^{\sigma_{j+1}}$ can be recursively constructed starting from $\Psi_{\vP}^{\sigma_0}\equiv \Omega_f$ with the help of the spectral projection $$\frac{1}{2\pi
  i}\oint_{\gamma_{j+1}}dz_{j+1}\frac{1}{H_{\vP}^{\sigma_{j+1}}-z_{j+1}}\,.$$
  More precisely,
\begin{eqnarray}
\Psi_{\vP}^{\sigma_{j+1}}&:=&\frac{1}{2\pi
  i}\oint_{\gamma_{j+1}}dz_{j+1}\frac{1}{H_{\vP}^{\sigma_{j+1}}-z_{j+1}}\Psi_{\vP}^{\sigma_{j}}\otimes\Omega_f\,\\
&= &\frac{1}{2\pi
  i}\sum_{n=0}^{\infty}\oint_{\gamma_{j+1}}dz_{j+1}\frac{1}{H_{\vP}^{\sigma_{j}}-z_{j+1}}[-\Delta H_{\vP}|^{\sigma_j}_{\sigma_{j+1}}\frac{1}{H_{\vP}^{\sigma_{j}}-z_{j+1}}]^n\,\Psi_{\vP}^{\sigma_{j}}\otimes\Omega_f \,,\quad\quad\quad\,\,\label{eq:II.13}
\end{eqnarray}
where $\gamma_{j+1}:=\{z_{j+1}\,\in\CC\,|\,|z_{j+1}-E_{\vP}^{\sigma_j}|=\mu\sigma_{j+1}\}$, with $\mu$ as in (\ref{eq:III.8bis}).
$\Psi_{\vP}^{\sigma_{j+1}}$ is the (unnormalized) ground state vector of
$H_{\vP}^{\sigma_{j+1}}|_{\cF_{\sigma}}$ for any $0\leq\sigma\leq \sigma_{j+1}$.
\end{itemize}
%
\begin{figure*}
\includegraphics[width=140mm,height=60mm]{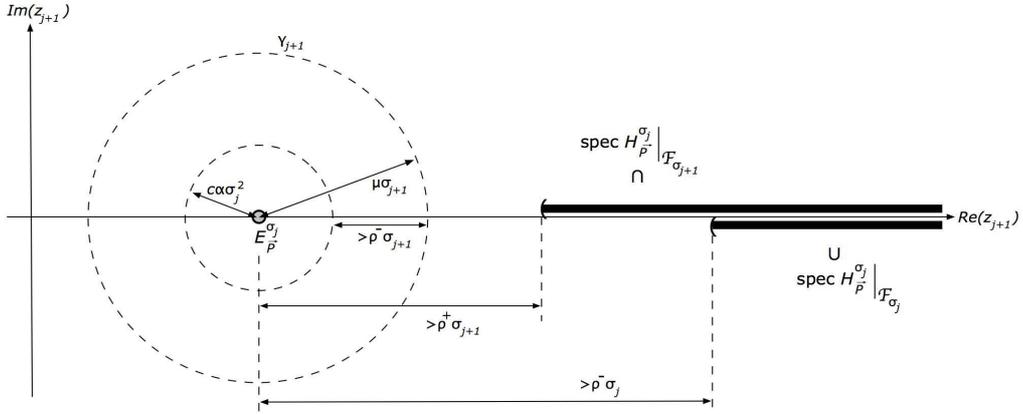}
\caption{The contour integral in the energy plane and the gaps.}
\end{figure*}
\subsection{Transformed Hamiltonians.}
In this section, we consider the (Bogoliubov-transformed) Hamiltonians
\begin{equation}
K_{\vP}^{\sigma_j}\,:=\,W_{\sigma_j}(\vnabla
E_{\vP}^{\sigma_j})H_{\vP}^{\sigma_j}W_{\sigma_j}^*(\vnabla
E_{\vP}^{\sigma_j})\,
\end{equation}
with ground state vectors $\Phi_{\vP}^{\sigma_j}$, $j=0,1,2,3,\dots$.
Some algebraic manipulations to express $K_{\vP}^{\sigma_j}$ in a
``canonical form" appear to represent a crucial step before
iterative perturbation theory can be applied to the sequence of these transformed Hamiltonians. In addition, some
intermediate Hamiltonians, denoted $\hat{K}_{\vP}^{\sigma_j}$,  must be introduced to arrive at the right kind of convergence estimates. 

\noindent
The same algebraic relations that are used to
obtain the ``canonical form" of $K_{\vP}^{\sigma_j}$ also play  an
important role in the proof of our main result concerning
the limiting behavior, as $\sigma\to 0$, of the second derivative of the ground
state energy $E_{\vP}^{\sigma}$. It is therefore useful to derive the
``canonical form" of $K_{\vP}^{\sigma_j}$ and the relevant algebraic identities in some detail.
\\
The Feynman-Hellman formula (which holds because $(H_{\vP}^{\sigma_j})_{\vP \in \cS}$ is
an analytic family of type A, and $E_{\vP}^{\sigma_j}$ is an isolated eigenvalue) yields the identity
\begin{equation}\label{eq-III.15}
\vnabla E_{\vP}^{\sigma_j}\,=\,\vP-\langle \vP^f-\alpha^{\frac{1}{2}}\vA^{\sigma_j}\rangle_{\psi_{\vP}^{\sigma_j}}\,,
\end{equation}
where,  given an operator $B$ and a vector $\psi$ in the domain of $B$, we use the notation
\begin{equation}\label{eq:III.16}
\langle
B \rangle_{\psi}\,:=\,\frac{\langle
  \psi\,,\,B\,\psi \rangle}{\langle \psi\,,\,\psi \rangle}\,.
\end{equation}

\noindent
As stated in \cite{ChenFroehlichPizzo2},  for $\alpha$ small enough,  $$\sup_{\vP\in \mathcal{S}}|\vnabla E_{\vP}^{\sigma_j}|<1 \quad \forall j \in \NN_0.$$

\noindent
We define
\begin{eqnarray}
\vec{\beta}^{\sigma_j}&:=&\vP^f-\alpha^{\frac{1}{2}}\vA^{\sigma_j}\label{eq-III.17}\\
\delta_{\vP}^{\sigma_j}(\hat{k})&:=&1-\hat{k}\cdot\vnabla E_{\vP}^{\sigma_j}, \quad\hat{k}:=\frac{\vk}{|\vk|}\,,\quad |\vnabla E_{\vP}^{\sigma_j}|<1 \label{eq-III.18}\\
c^{\,*}_{\vk,\lambda}&:=&b^{\,*}_{\vk,\lambda}+\alpha^{\frac{1}{2}}\frac{\vnabla
  E_{\vP}^{\sigma_j}\cdot\vec{\epsilon}^{\,\,*}_{\vk,\lambda}}{|\vk|^{\frac{3}{2}}\delta_{\vP}^{\sigma_j}(\hat{k})}\\
c_{\vk,\lambda}&:=&b_{\vk,\lambda}+\alpha^{\frac{1}{2}}\frac{\vnabla
  E_{\vP}^{\sigma_j}\cdot\vec{\epsilon}_{\vk,\lambda}}{|\vk|^{\frac{3}{2}}\delta_{\vP}^{\sigma_j}(\hat{k})}\,.
\end{eqnarray}
We rewrite $H_{\vP}^{\sigma_j}$ as 
\begin{equation}
H_{\vP}^{\sigma_j}\,=\,\frac{(\vP-\vec{\beta}^{\sigma_j})^2}{2}+H^f\,,
\end{equation}
and, using (\ref{eq-III.15}) and (\ref{eq-III.17}),
\begin{equation}
\vP\,=\,\vnabla
  E_{\vP}^{\sigma_j}+\langle \vec{\beta}^{\sigma_j}\rangle_{\psi_{\vP}^{\sigma_j}}\,.
\end{equation}
We then obtain
\begin{eqnarray}
H_{\vP}^{\sigma_j}&=&\frac{\vP^2}{2}-(\vnabla
  E_{\vP}^{\sigma_j}+\langle
  \vec{\beta}^{\sigma_j}\rangle_{\psi_{\vP}^{\sigma_j}})\cdot
  \vec{\beta}^{\sigma_j}+\frac{\vec{\beta}^{{\sigma_j}^2}}{2}+H^f\\
&=&\frac{\vP^2}{2}+\frac{\vec{\beta}^{{\sigma_j}^2}}{2}-\langle
  \vec{\beta}^{\sigma_j}\rangle_{\psi_{\vP}^{\sigma_j}}\cdot
  \vec{\beta}^{\sigma_j}\\
&
&+\sum_{\lambda}\int_{\RR^3\setminus(\cB_{\Lambda}\setminus\cB_{\sigma_j})}|\vk|\delta_{\vP}^{\sigma_j}(\hat{k})b^{\,*}_{\vk,\lambda}b_{\vk,\lambda}d^3k\\
&
&+\sum_{\lambda}\int_{\cB_{\Lambda}\setminus\cB_{\sigma_j}}|\vk|\delta_{\vP}^{\sigma_j}(\hat{k})c^{\,*}_{\vk,\lambda}c_{\vk,\lambda}d^3k\\
& &-\alpha\sum_{\lambda}\int_{\cB_{\Lambda}\setminus\cB_{\sigma_j}}|\vk|\delta_{\vP}^{\sigma_j}(\hat{k})\frac{\vnabla
  E_{\vP}^{\sigma_j}\cdot\vec{\epsilon}_{\vk,\lambda}}{|\vk|^{\frac{3}{2}}\delta_{\vP}^{\sigma_j}(\hat{k})}\frac{\vnabla
  E_{\vP}^{\sigma_j}\cdot\vec{\epsilon}^{\,\,*}_{\vk,\lambda}}{|\vk|^{\frac{3}{2}}\delta_{\vP}^{\sigma_j}(\hat{k})}d^3k\,.\quad\quad
\end{eqnarray}
Adding and subtracting $1/2\,\langle
  \vec{\beta}^{\sigma_j}\rangle_{\psi_{\vP}^{\sigma_j}}^2$, one finds that
\begin{eqnarray}\label{eq:II.28bis}
H_{\vP}^{\sigma_j}&=&\frac{\vP^2}{2}-\frac{\langle
  \vec{\beta}^{\sigma_j}\rangle_{\psi_{\vP}^{\sigma_j}}^2}{2}+\frac{(\vec{\beta}^{\sigma_j}-\langle
  \vec{\beta}^{\sigma_j}\rangle_{\psi_{\vP}^{\sigma_j}})^2}{2}\\
&
&+\sum_{\lambda}\int_{\RR^3\setminus(\cB_{\Lambda}\setminus\cB_{\sigma_j})}|\vk|\delta_{\vP}^{\sigma_j}(\hat{k})b^{\,*}_{\vk,\lambda}b_{\vk,\lambda}d^3k\\
&
&+\sum_{\lambda}\int_{\cB_{\Lambda}\setminus\cB_{\sigma_j}}|\vk|\delta_{\vP}^{\sigma_j}(\hat{k})c^{\,*}_{\vk,\lambda}c_{\vk,\lambda}d^3k\\
& &-\alpha\sum_{\lambda}\int_{\cB_{\Lambda}\setminus\cB_{\sigma_j}}|\vk|\delta_{\vP}^{\sigma_j}(\hat{k})\frac{\vnabla
  E_{\vP}^{\sigma_j}\cdot\vec{\epsilon}_{\vk,\lambda}}{|\vk|^{\frac{3}{2}}\delta_{\vP}^{\sigma_j}(\hat{k})}\frac{\vnabla
  E_{\vP}^{\sigma_j}\cdot\vec{\epsilon}^{\,\,*}_{\vk,\lambda}}{|\vk|^{\frac{3}{2}}\delta_{\vP}^{\sigma_j}(\hat{k})}d^3k\,.\quad\quad\label{eq:II.31bis}
\end{eqnarray}
Next, we implement the Bogoliubov transformation
\begin{eqnarray}
b^{\,*}_{\vk,\lambda}&\rightarrow&W_{\sigma_j}(\vnabla E_{\vP}^{\sigma_j})b^{\,*}_{\vk,\lambda}W_{\sigma_j}^*(\vnabla E_{\vP}^{\sigma_j})\,=\,b^{\,*}_{\vk,\lambda}-\alpha^{\frac{1}{2}}\frac{\vnabla
  E_{\vP}^{\sigma_j}\cdot\vec{\epsilon}^{\,\,*}_{\vk,\lambda}}{|\vk|^{\frac{3}{2}}\delta_{\vP}^{\sigma_j}(\hat{k})} \, ,\\
b_{\vk,\lambda}&\rightarrow&W_{\sigma_j}(\vnabla E_{\vP}^{\sigma_j})b_{\vk,\lambda}W_{\sigma_j}^*(\vnabla E_{\vP}^{\sigma_j})\,=\,b_{\vk,\lambda}-\alpha^{\frac{1}{2}}\frac{\vnabla
  E_{\vP}^{\sigma_j}\cdot\vec{\epsilon}_{\vk,\lambda}}{|\vk|^{\frac{3}{2}}\delta_{\vP}^{\sigma_j}(\hat{k})}\,,
\end{eqnarray}
for $\vk\in \cB_{\Lambda}\setminus\cB_{\sigma_j}$, where $W_{\sigma_j}(\vnabla
E_{\vP}^{\sigma_j})$ is defined in (\ref{eq:II.3}). It is evident that $W_{\sigma_j}$ acts as the identity on
$\cF^b(L^2(\cB_{\sigma_j}\times \ZZ_2))$ and on
$\cF^b(L^2((\RR^3\setminus\cB_{\Lambda})\times \ZZ_2))$. 

\noindent
We define
the vector operators
\begin{equation}\label{eq:II.34bis}
\vec{\Pi}_{\vP}^{\sigma_j}\,:=\,W_{\sigma_j}(\vnabla
E_{\vP}^{\sigma_j})\vec{\beta}^{\sigma_j}W_{\sigma_j}^*(\vnabla
E_{\vP}^{\sigma_j})\\
-\langle W_{\sigma_j}(\vnabla
E_{\vP}^{\sigma_j})\vec{\beta}^{\sigma_j}W_{\sigma_j}^*(\vnabla
E_{\vP}^{\sigma_j}) \rangle_{\Omega_f}\,,
\end{equation}
noting that, by (\ref{eq-III.15}), (\ref{eq-III.17}), and (\ref{eq:II.34bis})
\begin{eqnarray}\label{eq:III.35bis}
\langle \vec{\beta}^{\sigma_j} \rangle_{\psi_{\vP}^{\sigma_j}}&=&\vP-\vnabla
E_{\vP}^{\sigma_j}\\
&= &\frac{\langle
  \Phi_{\vP}^{\sigma_j}\,,\,\vec{\Pi}_{\vP}^{\sigma_j}\Phi_{\vP}^{\sigma_j}\rangle}{\langle \Phi_{\vP}^{\sigma_j}\,,\,\Phi_{\vP}^{\sigma_j}\rangle}+\langle W_{\sigma_j}(\vnabla
E_{\vP}^{\sigma_j})\vec{\beta}^{\sigma_j}W_{\sigma_j}^*(\vnabla
E_{\vP}^{\sigma_j}) \rangle_{\Omega_f}\,, \label{eq:III.36bis}
\end{eqnarray}
where $\Phi_{\vP}^{\sigma_j}$ is the ground state of the Bogoliubov-transformed
Hamiltonian
\begin{equation}
K_{\vP}^{\sigma_j}\,:=\,W_{\sigma_j}(\vnabla
E_{\vP}^{\sigma_j})H_{\vP}^{\sigma_j}W_{\sigma_j}^*(\vnabla
E_{\vP}^{\sigma_j})\,.
\end{equation}
It is easy to see that
\begin{equation}\label{eq:III.38}
W_{\sigma_j}(\vnabla
E_{\vP}^{\sigma_j})\vec{\beta}^{\sigma_j}W_{\sigma_j}^*(\vnabla
E_{\vP}^{\sigma_j})-\langle \vec{\beta}^{\sigma_j} \rangle_{
    \Psi_{\vP}^{\sigma_j}}=\vec{\Pi}_{\vP}^{\sigma_j}-\langle
  \vec{\Pi}_{\vP}^{\sigma_j} \rangle_{\Phi_{\vP}^{\sigma_j}}\,.
\end{equation}
After inspecting straightforward operator domain questions (see \cite{ChenFroehlichPizzo2}), the ``canonical form" of $K_{\vP}^{\sigma_j}$ is given by
\begin{equation}\label{eq:III.39}
K_{\vP}^{\sigma_j}\,=\,\frac{(\vec{\Gamma}^{\sigma_j}_{\vP})^2}{2}+\sum_{\lambda}\int_{\RR^3}|\vk|\delta_{\vP}^{\sigma_j}(\hat{k})b^*_{\vk,\lambda}b_{\vk,\lambda}d^3k+\cE_{\vP}^{\sigma_j}\,,
\end{equation}
where
\begin{equation}\label{eq:II.40}
\vec{\Gamma}^{\sigma_j}_{\vP}\,:=\,\vec{\Pi}_{\vP}^{\sigma_j}-\langle
  \vec{\Pi}_{\vP}^{\sigma_j} \rangle_{\Phi_{\vP}^{\sigma_j}}\,,
\end{equation}
so that
\begin{equation}
\langle \vec{\Gamma}^{\sigma_j}_{\vP}\rangle_{\Phi_{\vP}^{\sigma_j}}=0\,,
\end{equation}
and 
\begin{eqnarray}
\cE^{\sigma_j}_{\vP}&:=&\frac{\vP^2}{2}-\frac{(\vP-\vnabla
E_{\vP}^{\sigma_j})^2}{2} \label{eq-II.42}\\
&&-\alpha\sum_{\lambda}\int_{\cB_{\Lambda}\setminus\cB_{\sigma_j}}|\vk|\delta_{\vP}^{\sigma_j}(\hat{k})\,\frac{\vnabla
  E_{\vP}^{\sigma_j}\cdot\vec{\epsilon}_{\vk,\lambda}}{|\vk|^{\frac{3}{2}}\delta_{\vP}^{\sigma_j}(\hat{k})}\frac{\vnabla
  E_{\vP}^{\sigma_j}\cdot\vec{\epsilon}^{\,\,*}_{\vk,\lambda}}{|\vk|^{\frac{3}{2}}\delta_{\vP}^{\sigma_j}(\hat{k})}d^3k\,.\nonumber
\end{eqnarray}
Eqs. (\ref{eq:III.38}), (\ref{eq:III.39})  follow by using that
\begin{eqnarray}
W_{\sigma_j}(\vnabla
E_{\vP}^{\sigma_j})c^{\,*}_{\vk,\lambda}W_{\sigma_j}^*(\vnabla
E_{\vP}^{\sigma_j})&=&b^{\,*}_{\vk,\lambda}\,,\\
W_{\sigma_j}(\vnabla
E_{\vP}^{\sigma_j})c_{\vk,\lambda}W_{\sigma_j}^*(\vnabla
E_{\vP}^{\sigma_j})&=&b_{\vk,\lambda}\,,
\end{eqnarray}
for $\vk\in\cB_{\Lambda}\setminus\cB_{\sigma_j}$.
\\
An intermediate Hamiltonian, $\hat{K}_{\vP}^{\sigma_{j+1}}$,  is defined by
\begin{equation}
\hat{K}_{\vP}^{\sigma_{j+1}}\,:=\,W_{\sigma_{j+1}}(\vnabla
E_{\vP}^{\sigma_j})H^{\sigma_{j+1}}_{\vP}W_{\sigma_{j+1}}^*(\vnabla
E_{\vP}^{\sigma_j})\,,
\end{equation}
where
\begin{equation}
W_{\sigma_{j+1}}(\vnabla E_{\vP}^{\sigma_j})\,:=\,\exp\big(\alpha^{\frac{1}{2}}\sum_{\lambda}\int_{\cB_{\Lambda}\setminus\cB_{\sigma_{j+1}}}d^3k\frac{\vnabla
  E_{\vP}^{\sigma_j}}{|\vk|^{\frac{3}{2}}\delta_{\vP}^{\sigma_j}(\hat{k})}\cdot(\vec{\epsilon}_{\vk,\lambda}b^*_{\vk,\lambda}-h.c.)\big)\,.
\end{equation}
We decompose $\hat{K}_{\vP}^{\sigma_{j+1}}$ into several different terms, similarly as $K_{\vP}^{\sigma_j}$. We recall that
\begin{equation}
H_{\vP}^{\sigma_{j+1}}\,=\,\frac{(\vP-\vec{\beta}^{\sigma_{j+1}})^2}{2}+H^f\,,
\end{equation}
and, by (\ref{eq:III.35bis}), 
\begin{equation}
\vP\,=\,\vnabla
  E_{\vP}^{\sigma_j}+\langle \vec{\beta}^{\sigma_j}\rangle_{\psi_{\vP}^{\sigma_j}}\,.
\end{equation}
It follows that (see also (\ref{eq:II.28bis})-(\ref{eq:II.31bis}))
\begin{eqnarray}
H_{\vP}^{\sigma_{j+1}}&=&\frac{\vP^2}{2}-(\vnabla
  E_{\vP}^{\sigma_j}+\langle
  \vec{\beta}^{\sigma_j}\rangle_{\psi_{\vP}^{\sigma_j}})\cdot
  \vec{\beta}^{\sigma_{j+1}}+\frac{\vec{\beta}^{{\sigma_{j+1}}^2}}{2}+H^f\\
&=&\frac{\vP^2}{2}+\frac{\vec{\beta}^{{\sigma_{j+1}}^2}}{2}-\langle
  \vec{\beta}^{\sigma_j}\rangle_{\psi_{\vP}^{\sigma_j}}\cdot
  \vec{\beta}^{\sigma_{j+1}}\\
&
&+\sum_{\lambda}\int_{\RR^3\setminus(\cB_{\Lambda}\setminus\cB_{\sigma_{j+1}})}|\vk|\delta_{\vP}^{\sigma_j}(\hat{k})b^{\,*}_{\vk,\lambda}b_{\vk,\lambda}d^3k\\
&
&+\sum_{\lambda}\int_{\cB_{\Lambda}\setminus\cB_{\sigma_{j+1}}}|\vk|\delta_{\vP}^{\sigma_j}(\hat{k})c^{\,*}_{\vk,\lambda}c_{\vk,\lambda}d^3k\\
& &-\alpha\sum_{\lambda}\int_{\cB_{\Lambda}\setminus\cB_{\sigma_{j+1}}}|\vk|\delta_{\vP}^{\sigma_j}(\hat{k})\frac{\vnabla
  E_{\vP}^{\sigma_j}\cdot\vec{\epsilon}_{\vk,\lambda}}{|\vk|^{\frac{3}{2}}\delta_{\vP}^{\sigma_j}(\hat{k})}\frac{\vnabla
  E_{\vP}^{\sigma_j}\cdot\vec{\epsilon}^{\,\,*}_{\vk,\lambda}}{|\vk|^{\frac{3}{2}}\delta_{\vP}^{\sigma_j}(\hat{k})}d^3k\,.\quad\quad
\end{eqnarray}
We now add and subtract $1/2 \, \langle
  \vec{\beta}^{\sigma_j}\rangle_{\psi_{\vP}^{\sigma_j}}^2$  and conjugate the resulting operator with the unitary operator $W_{\sigma_{j+1}}(\vnabla
  E_{\vP}^{\sigma_j})$.  
   After inspecting straightforward operator domain questions (see \cite{ChenFroehlichPizzo2}), we find that
\begin{eqnarray}\label{eq:II.54bis}
\hat{K}_{\vP}^{\sigma_{j+1}}&=&\frac{(\vec{\Gamma}^{\sigma_j}_{\vP}+\vec{\cL}_{\sigma_{j+1}}^{\sigma_j}+\vec{\cI}_{\sigma_{j+1}}^{\sigma_j})^2}{2}\\
& &+\sum_{\lambda}\int_{\RR^3}|\vk|\delta_{\vP}^{\sigma_j}(\hat{k})b^{\,*}_{\vk,\lambda}b_{\vk,\lambda}d^3k+\hat{\cE}_{\vP}^{\sigma_{j+1}}\,,
\end{eqnarray}
where
\begin{eqnarray}
\vec{\cL}_{\sigma_{j+1}}^{\sigma_j}&:=&-\alpha^{\frac{1}{2}}\sum_{\lambda}\int_{\cB_{\sigma_j}\setminus\cB_{\sigma_{j+1}}}\vk\,\frac{\vnabla
  E_{\vP}^{\sigma_j}\cdot\vec{\epsilon}^{\,\,*}_{\vk,\lambda}b_{\vk,\lambda}+h.c.}{|\vk|^{\frac{3}{2}}\delta_{\vP}^{\sigma_j}(\hat{k})}d^3k\,,\label{eq:III.57}\\
  & &-\alpha^{\frac{1}{2}}\vec{A}|_{\sigma_{j+1}}^{\sigma_j}\\
 \vec{\cI}_{\sigma_{j+1}}^{\sigma_j}&:=&\alpha\sum_{\lambda}\int_{\cB_{\sigma_j}\setminus\cB_{\sigma_{j+1}}}\vk\,\frac{\vnabla
  E_{\vP}^{\sigma_j}\cdot\vec{\epsilon}_{\vk,\lambda}}{|\vk|^{\frac{3}{2}}\delta_{\vP}^{\sigma_j}(\hat{k})}\frac{\vnabla
  E_{\vP}^{\sigma_j}\cdot\vec{\epsilon}^{\,\,*}_{\vk,\lambda}}{|\vk|^{\frac{3}{2}}\delta_{\vP}^{\sigma_j}(\hat{k})}d^3k\,,\\
 & &+\alpha\sum_{\lambda}\int_{\cB_{\sigma_j}\setminus\cB_{\sigma_{j+1}}}\,[\vec{\epsilon}_{\vk,\lambda}\frac{\vnabla
 E_{\vP}^{\sigma_j}\cdot\vec{\epsilon}^{\,\,*}_{\vk,\lambda}}{|\vk|^{\frac{3}{2}}\delta_{\vP}^{\sigma_j}(\hat{k})}+h.c.]\frac{d^3k}{\sqrt{|\vk|}}\nonumber\\
 \hat{\cE}_{\vP}^{\sigma_{j+1}}&:= & \frac{\vP^2}{2}-\frac{(\vP-\vnabla
E_{\vP}^{\sigma_j})^2}{2}\\
&&-\alpha\sum_{\lambda}\int_{\cB_{\Lambda}\setminus\cB_{\sigma_{j+1}}}|\vk|\delta_{\vP}^{\sigma_j}(\hat{k})\,\frac{\vnabla E_{\vP}^{\sigma_j}\cdot\vec{\epsilon}_{\vk,\lambda}}{|\vk|^{\frac{3}{2}}\delta_{\vP}^{\sigma_j}(\hat{k})}\frac{\vnabla E_{\vP}^{\sigma_j}\cdot\vec{\epsilon}^{\,\,*}_{\vk,\lambda}}{|\vk|^{\frac{3}{2}}\delta_{\vP}^{\sigma_j}(\hat{k})}d^3k\,.\nonumber
\end{eqnarray}
We also define the operators
\begin{equation}\label{eq:III.61}
\hat{\vec{\Pi}}_{\vP}^{\sigma_j}\,:=\,W_{\sigma_{j}}(\vnabla E_{\vP}^{\sigma_{j-1}})W_{\sigma_{j}}^{*}(\vnabla E_{\vP}^{\sigma_{j}})\vec{\Pi}_{\vP}^{\sigma_j}W_{\sigma_{j}}(\vnabla E_{\vP}^{\sigma_{j}})W_{\sigma_{j}}^{*}(\vnabla E_{\vP}^{\sigma_{j-1}})\,,
\end{equation}
and
\begin{equation}\label{eq:III.62}
\hat{\vec{\Gamma}}^{\sigma_j}_{\vP}\,:=\,\hat{\vec{\Pi}}_{\vP}^{\sigma_j}-\langle
 \hat{ \vec{\Pi}}_{\vP}^{\sigma_j} \rangle_{\hat{\Phi}_{\vP}^{\sigma_j}}\,,
\end{equation}
which are used in the proofs of convergence of the ground state vectors. Here,
$\hat{\Phi}_{\vP}^{\sigma_j}$ denotes the ground state vector of the
Hamiltonian $$ \hat{K}_{\vP}^{\sigma_j}:=W_{\sigma_{j}}(\vnabla
E_{\vP}^{\sigma_{j-1}})W_{\sigma_{j}}^{*}(\vnabla E_{\vP}^{\sigma_{j}})K_{\vP}^{\sigma_j}W_{\sigma_{j}}(\vnabla E_{\vP}^{\sigma_{j}})W_{\sigma_{j}}^{*}(\vnabla
E_{\vP}^{\sigma_{j-1}})\,.$$
Notice that 
\begin{equation}\label{eq:identity}
\hat{\vec{\Gamma}}^{\sigma_j}_{\vP}=W_{\sigma_{j}}(\vnabla E_{\vP}^{\sigma_{j-1}})W_{\sigma_{j}}^{*}(\vnabla E_{\vP}^{\sigma_{j}})\vec{\Gamma}_{\vP}^{\sigma_j}W_{\sigma_{j}}(\vnabla E_{\vP}^{\sigma_{j}})W_{\sigma_{j}}^{*}(\vnabla E_{\vP}^{\sigma_{j-1}})\,.
\end{equation}
An important identity used in
\cite{ChenFroehlichPizzo2} and in the sequel of the present paper is ($j\geq 1$)
\begin{eqnarray}\label{eq:III.63}
\hat{\vec{\Gamma}}^{\sigma_j}_{\vP}-\vec{\Gamma}^{\sigma_{j-1}}_{\vP}&=&\vnabla
E_{\vP}^{\sigma_j}-\vnabla
E_{\vP}^{\sigma_{j-1}}+\vec{\cL}^{\sigma_{j-1}}_{\sigma_j}\\
& &+\alpha\sum_{\lambda}\int_{\cB_{\sigma_{j-1}}\setminus\cB_{\sigma_{j}}}\vk\,\frac{\vnabla
  E_{\vP}^{\sigma_{j-1}}\cdot\vec{\epsilon}_{\vk,\lambda}}{|\vk|^{\frac{3}{2}}\delta_{\vP}^{\sigma_{j-1}}(\hat{k})}\frac{\vnabla
  E_{\vP}^{\sigma_{j-1}}\cdot\vec{\epsilon}^{\,*}_{\vk,\lambda}}{|\vk|^{\frac{3}{2}}\delta_{\vP}^{\sigma_{j-1}}(\hat{k})}d^3k \nonumber\\
 & &+\alpha\sum_{\lambda}\int_{\cB_{\sigma_{j-1}}\setminus\cB_{\sigma_{j}}}\,[\vec{\epsilon}_{\vk,\lambda}\frac{\vnabla
 E_{\vP}^{\sigma_{j-1}}\cdot\vec{\epsilon}^{\,*}_{\vk,\lambda}}{|\vk|^{\frac{3}{2}}\delta_{\vP}^{\sigma_{j-1}}(\hat{k})}+h.c.]\frac{d^3k}{\sqrt{|\vk|}}\,.\nonumber
\end{eqnarray}
Eq. (\ref{eq:III.63}) can be derived using (\ref{eq:II.34bis}), (\ref{eq:III.36bis}),
(\ref{eq:III.38}), (\ref{eq:II.40}), (\ref{eq:III.61}), and  (\ref{eq:III.62}).

\subsection{Convergence of the sequence $\{\Phi_{\vP}^{\sigma_j}\}_{j=0}^{\infty}$.
}
To pass from momentum scale $j$ to $j+1$, we proceed in two steps: First, we construct an intermediate vector, $\hat{\Phi}_{\vP}^{\sigma_{j+1}}$,  defined by 
\begin{equation}\label{eq:III.64}
\hat{\Phi}_{\vP}^{\sigma_{j+1}}\,:=\,\sum_{n=0}^{\infty}\frac{1}{2\pi i}\oint_{\gamma_{j+1}}dz_{j+1}\frac{1}{K_{\vP}^{\sigma_j}-z_{j+1}}[-\Delta K_{\vP}|_{\sigma_{j+1}}^{\sigma_j}\frac{1}{K_{\vP}^{\sigma_j}-z_{j+1}}]^n\Phi_{\vP}^{\sigma_{j}}\,,
\end{equation}
where 
\begin{eqnarray}
\Delta K_{\vP}|_{\sigma_{j+1}}^{\sigma_j}&:=& \hat{K}_{\vP}^{\sigma_{j+1}}-\hat{\cE}_{\vP}^{\sigma_{j+1}}+\cE_{\vP}^{\sigma_j}- K_{\vP}^{\sigma_j}\\
&= &\frac{1}{2}\big(\vec{\Gamma}_{\vP}^{\sigma_j}\cdot(\vec{\cL}_{\sigma_{j+1}}^{\sigma_j}+ \vec{\cI}_{\sigma_{j+1}}^{\sigma_j})+h.c.\big)+\frac{1}{2}(\vec{\cL}_{\sigma_{j+1}}^{\sigma_j}+ \vec{\cI}_{\sigma_{j+1}}^{\sigma_j})^2\,.\label{eq:III.53}\quad\quad\quad
\end{eqnarray}
Subsequently, we construct $\Phi_{\vP}^{\sigma_{j+1}}$ by setting
\begin{equation}\label{eq:II.bis.67}
\Phi_{\vP}^{\sigma_{j+1}}\,:=\,W_{\sigma_{j+1}}(\vnabla E_{\vP}^{\sigma_{j+1}})W_{\sigma_{j+1}}^{*}(\vnabla E_{\vP}^{\sigma_{j}})\hat{\Phi}_{\vP}^{\sigma_{j+1}}\,.
\end{equation}
The series in (\ref{eq:III.64}) is term-wise well-defined and converges
strongly to a non-zero vector, provided $\alpha$ is small enough (\emph{independently}
of $j$). The proof of this claim is based on  operator-norm estimates of the type used
in controlling the Neumann expansion in (\ref{eq:II.13}),
which requires an estimate of the spectral gap that follows from the unitarity of $W_{\sigma_{j}}(\vnabla E_{\vP}^{\sigma_{j}})$ and Result 1) described after Eq. (\ref{eq:II.10}).

\noindent
A key point in our proof  of convergence of the sequence
$\{\Phi_{\vP}^{\sigma_j}\}$ is to show that the term
\begin{equation}
\vec{\Gamma}_{\vP}^{\sigma_j}\cdot(\vec{\cL}_{\sigma_{j+1}}^{\sigma_j}+\vec{\cI}_{\sigma_{j+1}}^{\sigma_j})+h.c.
\end{equation}
appearing in (\ref{eq:III.53}), which is superficially ``marginal" in the
infrared, by power counting, is in fact ``irrelevant" (using the terminology of renormalization group
theory). This is a consequence of the orthogonality condition 
\begin{equation}\label{eq:III.55}
\langle \Phi_{\vP}^{\sigma_{j}}\,,\,\vec{\Gamma}_{\vP}^{\sigma_j} \Phi_{\vP}^{\sigma_{j}}\rangle=0\,,
\end{equation}
which, when combined with an inductive argument, implies that
\begin{equation}
\|(\frac{1}{K_{\vP}^{\sigma_j}-z_{j+1}})^{\frac{1}{2}}\,[\vec{\Gamma}_{\vP}^{\sigma_j}\cdot(\vec{\cL}_{\sigma_{j+1}}^{\sigma_j\,(+)}+\vec{\cI}_{\sigma_{j+1}}^{\sigma_j})]\,(\frac{1}{K_{\vP}^{\sigma_j}-z_{j+1}})^{\frac{1}{2}}\,\Phi_{\vP}^{\sigma_{j}}\|
\end{equation}
(where $\vec{\cL}_{\sigma_{j+1}}^{\sigma_j\,(+)}$ stands for the part which
contains only photon creation operators) is of order $\cO(\epsilon^{\eta j})$,
for some $\eta >0$ specified in \cite{ChenFroehlichPizzo2}. In particular, this suffices to show that
\begin{equation}\label{eq:III.58}
\| \hat{\Phi}_{\vP}^{\sigma_{j+1}}-\Phi_{\vP}^{\sigma_{j}}\|\leq\cO(\epsilon^{\frac{j+1}{2}(1-\delta)})\,
\end{equation}
for any $0<\delta<1$ provided $\alpha$ is sufficiently small. Finally, in Theorem 3.1 of ref. \cite{ChenFroehlichPizzo2}, it is proven that there is a non-zero vector in the Hilbert space corresponding to $\lim_{j\to\infty}\Phi_{\vP}^{\sigma_{j}}$, and that the rate of convergence is at least $\cO(\sigma^{\frac{1}{2}(1-\delta)})$ for any $0<\delta<1$ provided $\alpha$ is sufficiently small.
\\

\noindent
{\bf{Remark}}

\noindent
In Theorem 3.1 of ref. \cite{ChenFroehlichPizzo2}, for $\lim_{j\to\infty}\Phi_{\vP}^{\sigma_{j}}$, the range of values of $\alpha$ such that the rate of convergence, $\cO(\sigma^{\frac{1}{2}(1-\delta)})$, holds is not claimed to be uniform in $\delta$. The stronger result obtained in the next section (see (\ref{eq:III.bis.2}) and (\ref{eq:III.34bis})) implies that this range (corresponding to the rate $\cO(\sigma^{\frac{1}{2}(1-\delta)})$) is actually $\delta$-independent.

\subsubsection{Key ingredients}\label{SectII.2}
To prove convergence of the
sequence $\{\Phi_{\vP}^{\sigma_j}\}$ of ground state vectors of the Hamiltonians $K_{\vP}^{\sigma_j}$,  some further conditions on $\alpha$, $\epsilon$, and $\mu$ (see (\ref{eq:III.8bis}), (\ref{eq:III.9})) are required, in particular an upper bound
on $\mu$ and an upper bound on $\epsilon$ strictly smaller than the ones
imposed by inequalities (\ref{eq:III.8bis}), (\ref{eq:III.9}); (for details, see Lemma
A.3 in \cite{ChenFroehlichPizzo2}). We note
that the more restrictive conditions on $\mu$  and $\epsilon$ imply new bounds
on $\rho^{-}$ and $\rho^{+}$. Moreover, $\epsilon$ must satisfy a lower bound
$\epsilon>C \alpha^{\frac{1}{2}}$, with a multiplicative constant $C>0$
sufficiently large. 

\noindent
Some key inequalities needed in our analysis of the convergence properties of $\{\Phi_{\vP}^{\sigma_j}\}$ are summarized below. They will be marked by the symbol $(\mathcal{B})$. In order to reach some important improvements in our estimates of the convergence rate of $\Phi_{\vP}^{\sigma_{j}}$, as $j\to\infty$ (discussed in the next section), a refined estimate is needed that is stated
in $(\mathcal{B}2)$, and a new inequality, see
$(\mathcal{B}5)$, (analogous to $(\mathcal{B}3)$ and $(\mathcal{B}4)$)  is required.
\begin{itemize}
\item \emph{Estimates on the shift of the ground state energy and its gradient}\\
There are constants $C_1$, $C_{2}'$ such that the following inequalities hold.\\
$(\mathcal{B}1)$
\begin{equation}
|E_{\vP}^{\sigma_j}-E_{\vP}^{\sigma_{j+1}}|\leq C_1\,\alpha\,\epsilon^j\,;
\end{equation}
see \cite{ChenFroehlichPizzo2}.\\
\item $(\mathcal{B}2)$
\begin{equation}\label{eq:III.73}
|\vnabla E_{\vP}^{\sigma_{j+1}}-\vnabla E_{\vP}^{\sigma_{j}}|\leq C'_2\big(\|\hat{\Phi}_{\vP}^{\sigma_{j+1}}-\Phi_{\vP}^{\sigma_{j}}\|+\alpha^{\frac{1}{4}}\epsilon^{j+1}\big)\,.
\end{equation}
This is an improvement over a corresponding estimate in  \cite{ChenFroehlichPizzo2}: It can be
proven \emph{after} the results stated in Theorem 3.1
in  \cite{ChenFroehlichPizzo2}, in particular the uniform bound from below on $\langle  \Phi_{\vP}^{\sigma_{j}},
\Phi_{\vP}^{\sigma_{j}}\rangle$,  $\langle  \Phi_{\vP}^{\sigma_{j}},
\Phi_{\vP}^{\sigma_{j}}\rangle >\frac{2}{3}$, and following the steps in the proof of Lemma A.2 in  \cite{ChenFroehlichPizzo2}.
\item \emph{Bounds relating expectations of operators to expectations of their
    absolute values}
\\
There are constants $C_3, C_4, C_5>1$ such that the following inequalities
hold.
\\
$(\mathcal{B}3)$ For $z_{j+1}\in\gamma_{j+1}$,
\begin{eqnarray}
& & \big\langle (\Gamma_{\vP}^{\sigma_j})^{i}\,\Phi_{\vP}^{\sigma_{j}}\,,\,\big|\frac{1}{K_{\vP}^{\sigma_j}-z_{j+1}}\big|(\Gamma_{\vP}^{\sigma_j})^{i}\,\Phi_{\vP}^{\sigma_{j}}\big \rangle \\
& &\leq C_3\big|\big\langle(\Gamma_{\vP}^{\sigma_j})^{i}\,\Phi_{\vP}^{\sigma_{j}}\,,\,\frac{1}{K_{\vP}^{\sigma_j}-z_{j+1}}(\Gamma_{\vP}^{\sigma_j})^{i}\,\Phi_{\vP}^{\sigma_{j}}\big\rangle\big|\,,
\end{eqnarray}
where $(\Gamma_{\vP}^{\sigma_j})^{i}$ is the $i^{th}$ component of $\vec{\Gamma}_{\vP}^{\sigma_j}$.

\noindent
$(\mathcal{B}4)$ For $z_{j+1}\in\gamma_{j+1},$
\begin{eqnarray}
& & \big\langle (\cL_{\sigma_{j+1}}^{\sigma_j\,(+)})^l\,(\Gamma_{\vP}^{\sigma_j})^{i}\,\Phi_{\vP}^{\sigma_{j}}\,,\,\big|\frac{1}{K_{\vP}^{\sigma_j}-z_{j+1}}\big|(\cL_{\sigma_{j+1}}^{\sigma_j\,(+)})^l\,(\Gamma_{\vP}^{\sigma_j})^{i}\,\Phi_{\vP}^{\sigma_{j}}\big \rangle \\
& &\leq C_4\big|\big\langle(\cL_{\sigma_{j+1}}^{\sigma_j\,(+)})^l\,(\Gamma_{\vP}^{\sigma_j})^{i}\,\Phi_{\vP}^{\sigma_{j}}\,,\,\frac{1}{K_{\vP}^{\sigma_j}-z_{j+1}}(\cL_{\sigma_{j+1}}^{\sigma_j\,(+)})^l\,(\Gamma_{\vP}^{\sigma_j})^{i}\,\Phi_{\vP}^{\sigma_{j}}\big\rangle\big|\,,\quad\quad\quad\quad
\end{eqnarray}
where $(\cL_{\sigma_{j+1}}^{\sigma_j\,(+)})^l$ is the $l^{th}$ component of $\vec{\cL}_{\sigma_{j+1}}^{\sigma_j\,(+)}$.

\noindent
$(\mathcal{B}5)$ For $z_{j+1}\in\gamma_{j+1},$
\begin{eqnarray}\label{eq:II.78}
& & \big\langle (\Gamma_{\vP}^{\sigma_j})^{i}\,\Phi_{\vP}^{\sigma_{j}}\,,\,\big|\frac{1}{K_{\vP}^{\sigma_j}-z_{j+1}}\big|^2(\Gamma_{\vP}^{\sigma_j})^{i}\,\Phi_{\vP}^{\sigma_{j}}\big \rangle \\
& &\leq C_5\big|\big\langle(\Gamma_{\vP}^{\sigma_j})^{i}\,\Phi_{\vP}^{\sigma_{j}}\,,\,(\frac{1}{K_{\vP}^{\sigma_j}-z_{j+1}})^2(\Gamma_{\vP}^{\sigma_j})^{i}\,\Phi_{\vP}^{\sigma_{j}}\big\rangle\big|\,.\nonumber
\end{eqnarray}
\end{itemize}

To prove $(\mathcal{B}3)$ and $(\mathcal{B}4)$, it suffices to
exploit the fact that the spectral support (with respect to
$K_{\vP}^{\sigma_j}$) of the two vectors
$(\Gamma_{\vP}^{\sigma_j})^{i}\,\Phi_{\vP}^{\sigma_{j}}$ and
$(\cL_{\sigma_{j+1}}^{\sigma_j\,(+)})^{l}(\Gamma_{\vP}^{\sigma_j})^{i}\,\Phi_{\vP}^{\sigma_{j}}$
is strictly above the ground state energy of $K^{\sigma_j}_{\vP}$, since they are both orthogonal to
the ground state, $\Phi_{\vP}^{\sigma_{j}}$, of this operator. In the proof of bound $(\mathcal{B}5)$, it
is also required that $\rho^->3\mu\epsilon$, as will be assumed in the following.
\\

\noindent
\emph{\bf{Remarks} }

\noindent
(1) The constants $C_1,\dots, C_5$ are independent of $\alpha, \epsilon, \mu$, and $j\in\NN$, provided that $\alpha, \epsilon$, and $ \mu$ are sufficiently small. 

\noindent
(2) For the convenience of the reader, we recapitulate the relations between the parameters entering the construction:
\begin{eqnarray}
& &0<\rho^-<\mu<\rho^+<1-C_{\alpha}<\frac{2}{3}\,,\label{eq:II.79bis}\\
& &0<\epsilon<\frac{\rho^-}{\rho^+}\,,\\
& &\epsilon>C \alpha^{1/2}\,, \label{eq:II.83b}\\
& &\rho^->3\mu\epsilon\,.\label{eq:II.82bisbis}
\end{eqnarray}
Moreover, we stress that the final result is a small coupling result, i.e., valid for
small values of $\alpha$, and that, for technical reasons, small values of the
parameters $\epsilon, \mu$ within the
constraints listed above (that imply more restrictive bounds on $ \rho^-, \rho^+$) are required.
\\

The crucial estimate for the bound on
$\hat{\Phi}_{\vP}^{\sigma_{j+1}}-\Phi_{\vP}^{\sigma_{j}}$ obtained in
\cite{ChenFroehlichPizzo2}  (see (\ref{eq:III.58})) is
\begin{equation}\label{eq:III.65}
\big|\big\langle (\Gamma_{\vP}^{\sigma_j})^{i}\,\Phi_{\vP}^{\sigma_{j}}\,,\,\frac{1}{K_{\vP}^{\sigma_j}-z_{j+1}}(\Gamma_{\vP}^{\sigma_j})^{i}\,\Phi_{\vP}^{\sigma_{j}}\big \rangle \big|\leq \frac{R_0}{\alpha \epsilon^{j\delta}}\,,
\end{equation}
where  $R_0$ is independent of $j$,  and $\delta$,  $0<\delta<1$, can be taken  arbitrarily small for $\alpha$ and $ \epsilon$
sufficiently small (depending on $\delta$).
This estimate will be improved in the next section. As a consequence, our estimate
of the convergence rate of $\{\Phi_{\vP}^{\sigma_{j}}\}$ will be
improved.  As a corollary, the second derivative of $E^{\sigma}_{\vP}$ is proven to converge,
as $\sigma\to0$.
\section{Improved estimate of the convergence rate of $\{\Phi_{\vP}^{\sigma}\}$, as $\sigma\to0$, and uniform bound  on the second derivative of $E^{\sigma}_{\vP}$.}\label{SectIII}
\resetequ

\noindent
Our arguments in Section III rely on the results previously proven in \cite{ChenFroehlichPizzo2} and described in Section II, which hold for $\alpha$ small enough. Therefore, in the following, we assume the constraints  (\ref{eq:II.79bis})-(\ref{eq:II.82bisbis}), and we make use of the estimates on the spectral gaps (see $(\mathcal{A}1)$  in Section \ref{SectII.0}) and of  the bounds  $(\mathcal{B}1)$-$(\mathcal{B}5)$ (see Section \ref{SectII.2}). 

\noindent
We also make use of the lower bounds
\begin{equation}\label{eq:III.3bis}
\langle  \hat{\Phi}_{\vP}^{\sigma_{j+1}},
\hat{\Phi}_{\vP}^{\sigma_{j+1}}\rangle\,,\,\langle  \Phi_{\vP}^{\sigma_{j}},
\Phi_{\vP}^{\sigma_{j}}\rangle >\frac{2}{3}
\end{equation}
uniformly in $j\in\NN_0$,  which appear in the proof of 
Theorem 3.1 of ref. \cite{ChenFroehlichPizzo2}. Assuming these bounds we can simplify the proof by induction in the theorem below.
\\

\begin{theorem}\label{Th.induction}
For $\alpha, \epsilon$ sufficiently small (depending on $\delta$), the inequality
\begin{equation} \label{eq:IV.1}
\big|\big\langle (\Gamma_{\vP}^{\sigma_j})^{i}\,\Phi_{\vP}^{\sigma_{j}}\,,\,\big(\frac{1}{K_{\vP}^{\sigma_j}-z_{j+1}}\big)^2(\Gamma_{\vP}^{\sigma_j})^{i}\,\Phi_{\vP}^{\sigma_{j}}\big \rangle \big|\leq \frac{\cR_0}{\alpha^{\frac{1}{2}}\, \epsilon^{2j\delta}}\,
\end{equation}
holds true where $0<\delta<1$ and $\cR_0$ is a constant independent of  $j\in\NN_0:=\NN\cup0$. Furthermore, for $\cR_0$ and $\alpha$ small enough,  inequality (\ref{eq:IV.1}) implies
\begin{equation}\label{eq:III.bis.2}
\|\hat{\Phi}_{\vP}^{\sigma_{j}}-\Phi_{\vP}^{\sigma_{j-1}}\|\leq\alpha^{\frac{1}{4}}\epsilon^{j(1-\delta)}\,.
\end{equation}
\end{theorem}

\noindent
\emph{Proof by induction.}

\begin{itemize}
\item \emph{\underline{Inductive hypothesis}} 

\noindent
We assume that, at scale  $j-1(\geq 0)$,
the  following estimate holds
\begin{equation}\label{eq:IV.3}
\big|\big\langle (\Gamma_{\vP}^{\sigma_{j-1}})^{i}\,\Phi_{\vP}^{\sigma_{j-1}}\,,\,\big(\frac{1}{K_{\vP}^{\sigma_{j-1}}-z_{j}}\big)^2(\Gamma_{\vP}^{\sigma_{j-1}})^{i}\,\Phi_{\vP}^{\sigma_{j-1}}\big \rangle \big| \leq \frac{\cR_0}{\alpha^{\frac{1}{2}}\, \epsilon^{2(j-1)\delta}}\,.
\end{equation}
This estimate readily implies that, for $\cR_0$ and $ \alpha$ small enough, but uniformly in $j$, 
\begin{eqnarray}\label{eq:III.70}
& &\|\hat{\Phi}_{\vP}^{\sigma_{j}}-\Phi_{\vP}^{\sigma_{j-1}}\|\,\\
& =&\,\|\sum_{n=1}^{\infty}\frac{1}{2\pi i}\oint_{\gamma_{j}}dz_{j}\frac{1}{K_{\vP}^{\sigma_{j-1}}-z_{j}}[-\Delta K_{\vP}|_{\sigma_{j}}^{\sigma_{j-1}}\frac{1}{K_{\vP}^{\sigma_{j-1}}-z_{j}}]^n\Phi_{\vP}^{\sigma_{j-1}}\|\,\nonumber\\
&\leq&\alpha^{\frac{1}{4}}\epsilon^{j(1-\delta)}\,.\label{eq:III.71}
\end{eqnarray}
An improved estimate on
$\|\hat{\Phi}_{\vP}^{\sigma_{j}}-\Phi_{\vP}^{\sigma_{j-1}}\|$ is based on the
following bounds:
\begin{itemize}
\item[i)]
\begin{equation}
\|\frac{1}{K_{\vP}^{\sigma_{j-1}}-z_{j}}\Delta
K_{\vP}|_{\sigma_{j}}^{\sigma_{j-1}}\Phi_{\vP}^{\sigma_{j-1}}\|\leq \cO(\cR_0^{\frac{1}{2}}\alpha^{\frac{1}{4}}\epsilon^{j(1-\delta)})\,,
\end{equation}
whose proof requires the use of the ``pull-through formula" (see, e.g., \cite{Schweber}), a Neumann expansion of the resolvent, the inequality in Eq. (\ref{eq:II.78}), and Eq. (\ref{eq:IV.3}); the reader can follow the similar steps used in Lemma A3 of ref.
\cite{ChenFroehlichPizzo2};
\item[ii)]
\begin{equation}
\|\frac{1}{K_{\vP}^{\sigma_{j-1}}-z_{j}}\Delta K_{\vP}|_{\sigma_{j}}^{\sigma_{j-1}}\|_{\cF_{\sigma_j}}\leq\cO(\alpha^{\frac{1}{2}})\,;
\end{equation}
this estimate can be derived from standard bounds and using the ``pull-through formula".
\end{itemize}

\item  \emph{\underline{Induction step from scale $j-1$ to scale $j$}} 

 By unitarity of $W_{\sigma_{j}}(\vnabla E_{\vP}^{\sigma_{j-1}})W_{\sigma_{j}}^*(\vnabla E_{\vP}^{\sigma_{j}})$, we have that 
\begin{eqnarray}
\lefteqn{\big|\big\langle (\Gamma_{\vP}^{\sigma_j})^{i}\,\Phi_{\vP}^{\sigma_{j}}\,,\,\big(\frac{1}{K_{\vP}^{\sigma_j}-z_{j+1}}\big)^2(\Gamma_{\vP}^{\sigma_j})^{i}\,\Phi_{\vP}^{\sigma_{j}}\big \rangle\big|}\nonumber\\
&= &\big|\big\langle (\hat{\Gamma}_{\vP}^{\sigma_j})^{i}\,\hat{\Phi}_{\vP}^{\sigma_{j}}\,,\,\big(\frac{1}{\hat{K}_{\vP}^{\sigma_j}-z_{j+1}}\big)^2(\hat{\Gamma}_{\vP}^{\sigma_j})^{i}\,\hat{\Phi}_{\vP}^{\sigma_{j}}\big \rangle \big|\,.
\end{eqnarray}
For $\alpha$ small enough and $\epsilon>C\,\alpha^{\frac{1}{2}}$, where $C>0$ is large enough, we may use $(\mathcal{B}1)$ to re-expand the resolvent and find that
\begin{eqnarray}
\lefteqn{\big|\big\langle (\hat{\Gamma}_{\vP}^{\sigma_j})^{i}\,\hat{\Phi}_{\vP}^{\sigma_{j}}\,,\,\big(\frac{1}{\hat{K}_{\vP}^{\sigma_j}-z_{j+1}}\big)^2(\hat{\Gamma}_{\vP}^{\sigma_j})^{i}\,\hat{\Phi}_{\vP}^{\sigma_{j}}\big \rangle \big|}\\
& &\leq 2\big|\big\langle (\hat{\Gamma}_{\vP}^{\sigma_j})^{i}\,\hat{\Phi}_{\vP}^{\sigma_{j}}\,,\,\big|\frac{1}{K_{\vP}^{\sigma_{j-1}}-z_{j+1}}\big|^2(\hat{\Gamma}_{\vP}^{\sigma_j})^{i}\,\hat{\Phi}_{\vP}^{\sigma_{j}}\big \rangle \big|\,.
\end{eqnarray}
It follows that 
\begin{eqnarray}
& &2\big|\big\langle (\hat{\Gamma}_{\vP}^{\sigma_j})^{i}\,\hat{\Phi}_{\vP}^{\sigma_{j}}\,,\,\big|\frac{1}{K_{\vP}^{\sigma_{j-1}}-z_{j+1}}\big|^2(\hat{\Gamma}_{\vP}^{\sigma_j})^{i}\,\hat{\Phi}_{\vP}^{\sigma_{j}}\big \rangle \big|\\
& &\leq 4\big \|\big|\frac{1}{K_{\vP}^{\sigma_{j-1}}-z_{j+1}}\big|\,((\hat{\Gamma}_{\vP}^{\sigma_j})^{i}\hat{\Phi}_{\vP}^{\sigma_{j}}-(\Gamma_{\vP}^{\sigma_{j-1}})^{i}\Phi_{\vP}^{\sigma_{j-1}})\big\|^2 \label{eq:III.79} \\
& & \quad+4\big|\big\langle (\Gamma_{\vP}^{\sigma_{j-1}})^{i}\,\Phi_{\vP}^{\sigma_{j-1}}\,,\,\big|\frac{1}{K_{\vP}^{\sigma_{j-1}}-z_{j+1}}\big|^2(\Gamma_{\vP}^{\sigma_{j-1}})^{i}\,\Phi_{\vP}^{\sigma_{j-1}}\big \rangle \big| \,.\,\,\,\label{eq:III.80}
\end{eqnarray}
Our recursion, combined with (\ref{eq:II.78}),  relates (\ref{eq:III.80}) to the initial expression in (\ref{eq:IV.1}), with $j$ replaced by $j-1$, while  (\ref{eq:III.79}) is a remainder term. Next we note that
\begin{eqnarray}
& & 4\big \|\big|\frac{1}{K_{\vP}^{\sigma_{j-1}}-z_{j+1}}\big|\,((\hat{\Gamma}_{\vP}^{\sigma_j})^{i}\hat{\Phi}_{\vP}^{\sigma_{j}}-(\Gamma_{\vP}^{\sigma_{j-1}})^{i}\Phi_{\vP}^{\sigma_{j-1}})\big\|^2\\
&\leq & 8\big \|\big|\frac{1}{K_{\vP}^{\sigma_{j-1}}-z_{j+1}}\big|\,((\hat{\Gamma}_{\vP}^{\sigma_{j}})^{i}\hat{\Phi}_{\vP}^{\sigma_{j}}-(\Gamma_{\vP}^{\sigma_{j-1}})^{i}\hat{\Phi}_{\vP}^{\sigma_{j}})\big\|^2\label{eq:III.82}\\
& &+8\big \|\big|\frac{1}{K_{\vP}^{\sigma_{j-1}}-z_{j+1}}\big|\,(\Gamma_{\vP}^{\sigma_{j-1}})^{i}(\hat{\Phi}_{\vP}^{\sigma_{j}}-\Phi_{\vP}^{\sigma_{j-1}})\big\|^2\label{eq:III.83}\\
&\leq &\frac{\mathcal{R}_1}{\epsilon^{2j\delta}}+\frac{\mathcal{R}_2}{\epsilon^{2j\delta}}\,.\label{eq:III.84}
\end{eqnarray}
Here $\mathcal{R}_1\leq\cO(\epsilon^{-2})$ and $\mathcal{R}_2\leq\cO(\epsilon^{-2})$ are constants independent of $\alpha$, $\mu$, and $j\in\NN$, provided that $\alpha$, $\mu$ are sufficiently small, and $\epsilon>C\alpha^{\frac{1}{2}}$. In detail:
\begin{itemize}
\item
Property ($\mathcal{B}4$) and the two norm-bounds
\begin{equation}
 \|\frac{1}{K_{\vP}^{\sigma_{j-1}}-z_{j+1}}(\Gamma_{\vP}^{\sigma_{j-1}})^{i}\|_{\cF_{\sigma_j}}\leq\cO(\epsilon^{-(j+1)})\quad,\quad\quad\|\hat{\Phi}_{\vP}^{\sigma_{j}}-\Phi_{\vP}^{\sigma_{j-1}}\|\leq\alpha^{\frac{1}{4}}\epsilon^{j(1-\delta)}
\end{equation}
(see (\ref{eq:III.70})) justify the step from (\ref{eq:III.83}) to
(\ref{eq:III.84});
\item
concerning the step from (\ref{eq:III.82}) to (\ref{eq:III.84}),
it is enough to consider
Eq.~(\ref{eq:III.63}) and the two bounds
\begin{equation}\label{eq:III.20b}
 \|(\cL^{\sigma_{j-1}}_{\sigma_j})^{i}\hat{\Phi}_{\vP}^{\sigma_j}\|\leq\cO(\alpha^{\frac{1}{2}}\epsilon^{j-1})\quad,\quad\|\hat{\Phi}_{\vP}^{\sigma_{j}}-\Phi_{\vP}^{\sigma_{j-1}}\|\leq\alpha^{\frac{1}{4}}\epsilon^{j(1-\delta)}\,.
\end{equation}
(hint: for the first inequality in (\ref{eq:III.20b}), use the expression in (\ref{eq:III.64}).)
\end{itemize}
To bound the term (\ref{eq:III.80}), we use $(\mathcal{B}5)$ and the key orthogonality property (\ref{eq:III.55}). For  $z_{j}\in\gamma_{j}$ and $z_{j+1}\in\gamma_{j+1}$, we find that for $\epsilon/\rho^{-}$ sufficiently small
\begin{eqnarray}
& &4\big|\big\langle (\Gamma_{\vP}^{\sigma_{j-1}})^{i}\,\Phi_{\vP}^{\sigma_{j-1}}\,,\,\big|\frac{1}{K_{\vP}^{\sigma_{j-1}}-z_{j+1}}\big|^2(\Gamma_{\vP}^{\sigma_{j-1}})^{i}\,\Phi_{\vP}^{\sigma_{j-1}}\big \rangle \big|\\
& &\leq 4C_5\big|\big\langle (\Gamma_{\vP}^{\sigma_{j-1}})^{i}\,\Phi_{\vP}^{\sigma_{j-1}}\,,\,\big(\frac{1}{K_{\vP}^{\sigma_{j-1}}-z_{j+1}}\big)^2(\Gamma_{\vP}^{\sigma_{j-1}})^{i}\,\Phi_{\vP}^{\sigma_{j-1}}\big \rangle \big|\quad\quad\quad \label{eq:III.87}\\
& &\leq 8C_5^2\big|\big\langle (\Gamma_{\vP}^{\sigma_{j-1}})^{i}\,\Phi_{\vP}^{\sigma_{j-1}}\,,\,\big(\frac{1}{K_{\vP}^{\sigma_{j-1}}-z_{j}}\big)^2(\Gamma_{\vP}^{\sigma_{j-1}})^{i}\,\Phi_{\vP}^{\sigma_{j-1}}\big \rangle \big|\,.\label{eq:III.88}
\end{eqnarray}
In passing from (\ref{eq:III.87}) to (\ref{eq:III.88}), we again use the constraint on the spectral support (with respect to $K_{\vP}^{\sigma_{j-1}}$) of the vector $(\Gamma_{\vP}^{\sigma_{j-1}})^{i}\,\Phi_{\vP}^{\sigma_{j-1}}$.

Assuming that the parameters $\epsilon$ and $\alpha$ are so small that the previous constraints are fulfilled and that
\begin{equation}\label{eq:III.89}
0<\cR_1+\cR_2\leq\,(1-8C_5^2\epsilon^{2\delta})\frac{\cR_0}{\alpha^{\frac{1}{2}}}\,,
\end{equation} 
we then conclude that
\begin{eqnarray}
& &\big|\big\langle (\Gamma_{\vP}^{\sigma_j})^{i}\,\Phi_{\vP}^{\sigma_{j}}\,,\,\big(\frac{1}{\hat{K}_{\vP}^{\sigma_j}-z_{j+1}}\big)^2(\Gamma_{\vP}^{\sigma_j})^{i}\,\Phi_{\vP}^{\sigma_{j}}\big \rangle \big|\\
&\leq &\frac{\mathcal{R}_1}{\epsilon^{2j\delta}}+\frac{\mathcal{R}_2}{\epsilon^{2j\delta}}\\
& &+8C_5^2\big|\big\langle (\Gamma_{\vP}^{\sigma_{j-1}})^{i}\,\Phi_{\vP}^{\sigma_{j-1}}\,,\,\big(\frac{1}{K_{\vP}^{\sigma_{j-1}}-z_{j}}\big)^2(\Gamma_{\vP}^{\sigma_{j-1}})^{i}\,\Phi_{\vP}^{\sigma_{j-1}}\big \rangle \big|\,\quad\\
&\leq&\frac{\cR_0}{\alpha^{\frac{1}{2}}\, \epsilon^{2j\delta}}\,.
\end{eqnarray}

\noindent
Notice that the bound in (\ref{eq:III.89}) induces a $\delta-$dependent constraint on the admissible values of  $\epsilon$ and, due to (\ref{eq:II.83b}), on $\alpha$.

\noindent
\item  \emph{\underline{The zeroth step in the induction}}

\noindent
Since 
\begin{equation}
(\Gamma_{\vP}^{\sigma_{0}})^{i}\equiv (\vP^f)^{i}\quad,\quad\Phi_{\vP}^{\sigma_{0}}\equiv\Omega_f\,,
\end{equation}
inequality (\ref{eq:IV.1}) is trivially fulfilled for $j=0$; thus
(\ref{eq:IV.1}) holds for all $j\in\NN_0$ and for $\cR_0$ \emph{arbitrarily} small, 
provided $\alpha$ is small enough. \QED
\end{itemize}

\noindent
As we explain below, an improved estimate of the rate of convergence of the sequence $\{\Phi_{\vec{P}}^{\sigma_{j}}\}_{j=0}^{\infty}$ follows from  the bound in (\ref{eq:III.bis.2}),  but we stress that only the estimates  in Eqs. (\ref{eq:IV.1}), (\ref{eq:III.bis.2}) will be used for the uniform bound on the second derivative of $E^{\sigma}_{\vP}$ in next section. 

\noindent
In fact, one can combine the bound in Eq. (\ref{eq:III.bis.2}) with the estimate
\begin{equation}\label{eq:III.30bis}
    \|\Phi_{\vec{P}}^{\sigma_{j}}-\widehat{\Phi}_{\vec{P}}^{\sigma_{j}}\|
    \, \leq \, C \, \alpha^{\frac{1}{2}}
    \,|\vnabla E_{\vec{P}}^{\sigma_{j-1}}-\vnabla E_{\vec{P}}^{\sigma_{j}}|\,|\ln(\epsilon^j)|\,,
\end{equation} 
where $C$ is independent of $\alpha$, $\epsilon$, $\mu$, and $j\in\NN$, provided that $\alpha$, $\epsilon$,  and $\mu$ are sufficiently small. The estimate in (\ref{eq:III.30bis}) is obtained starting from 
the definition in Eq. (\ref{eq:II.bis.67}) and using the soft photon bound
\begin{equation}\label{eq-II-9bisbis}
  	\| \, b_{\vk,\lambda}\Psi_{\vP}^{\sigma_j} \, \| \, \leq C \, \alpha^{1/2}
  	\, \frac{\one_{\sigma_j,\Lambda}(\vk)}{|\vk|^{3/2}}\,,\quad \one_{\sigma_j,\Lambda}(\vk):=\{\vk\,:\,\sigma_j<|\vk|\leq \Lambda\}\,,
\end{equation}
that follows from inequality (\ref{eq:II.10}) and the
identity
\begin{equation}\label{eq-II-9bis}
  b_{\vk,\lambda}\Psi_{\vP}^{\sigma_j}\,=\,-\,
  \alpha^{\frac{1}{2}}\,\frac{\one_{\sigma_j,\Lambda}(\vk)}{|\vk|^{\frac{1}{2}}}
  \, \frac{1}{H_{\vP-\vk}^{\sigma_j}+|\vk|-E_{\vP}^{\sigma_j}}
  \, \veps_{\vk,\lambda}\cdot\vnabla_{\vP}
  H_{\vP}^{\sigma_j}\,\Psi_{\vP}^{\sigma_j}\,,
\end{equation}
which is derived in \cite{ChFr} by using a
``pull-through argument".  

\noindent
By a standard procedure (see, e.g.,  \cite{Pizzo}), one obtains similar results for the
ground state vectors of the $\sigma$-dependent Hamiltonians $K_{\vP}^{\sigma}$,
for arbitrary $\sigma>0$. A precise statement concerning the rate of convergence is as follows:  The \emph{normalized} ground state vectors (that, with an abuse
of notation, we call $\Phi_{\vP}^{\sigma}$) 
\begin{equation}
\Phi_{\vP}^{\sigma}\,:=\frac{\frac{1}{2\pi
  i}\oint_{\gamma_{\sigma}}dz \frac{1}{K_{\vP}^{\sigma}-z}\Omega_f }{\|\frac{1}{2\pi
  i}\oint_{\gamma_{\sigma}}\frac{1}{K_{\vP}^{\sigma}-z}\Omega_f \|}\,,
 \end{equation}
where $\gamma_{\sigma}:=\{z\in\CC\,|\,|z-E_{\vP}^{\sigma}|=\frac{\rho^{-}}{2}\,\sigma\}$, converge strongly to a vector $\Phi_{\vP}$, as $\sigma\to0$,  with
\begin{equation}\label{eq:III.34bis}
\|\Phi_{\vP}^{\sigma}-\Phi_{\vP}\|\leq \cO(\alpha^{\frac{1}{4}}\,\big(\frac{\sigma}{\Lambda}\big)^{1-\delta})
\end{equation}
for any $0<\delta(<1)$, provided $\alpha$ is in an interval $(0,\alpha_{\delta})$  where $\alpha_{\delta}>0$, $\alpha_{\delta}\to 0$ as $\delta\to0$. The $\delta$-dependence of the interval $\alpha_{\delta}$ is an indirect consequence of the upper bound on $\epsilon$ that must be imposed through (\ref{eq:III.89}) to implement the proof by induction. The relations (\ref{eq:II.79bis})-(\ref{eq:II.82bisbis}) induce a $\delta$-dependence on the other parameters and in particular on $\alpha$.  Another $\delta$-type dependence of the estimated rate of convergence, $\cO(\big(\frac{\sigma}{\Lambda}\big)^{1-\delta})$, comes from the logarithmic term in (\ref{eq:III.30bis}). However, this does not spoil the uniformity in $\delta$ of the interval of admissible values of $\alpha$ but only affects the multiplicative constant on the R.H.S. of (\ref{eq:III.34bis}). Moreover, with further work, the estimate in Eq. (\ref{eq:III.30bis}) can be improved to remove the logarithmic term.
\subsection{Convergence of the second derivative of the ground state energy $E^{\sigma}_{\vP}$.}\label{SectIII.1}

Because of rotational symmetry we have that $ E^{\sigma}_{\vP}\equiv
E^{\sigma}_{|\vP|}$. Moreover,  $(H_{\vP}^{\sigma})_{\vP\in\cS}$ is an analytic
family of type A in $\vP\in\cS$, with an isolated eigenvalue $E^{\sigma}_{|\vP|}$. Thus, the second derivative $\frac{\partial^2E^{\sigma}_{|\vP|}}{(\partial |\vP|)^2}$ is well defined and
\begin{equation}\label{eq:IV.33a}
\frac{\partial^2E^{\sigma}_{|\vP|}}{(\partial
  |\vP|)^2}\,=\,\partial_i^2E^{\sigma}_{|\vP|}|_{\vP=P^{i}\hat{i}}\,,\quad i=1,2,3\,,
\end{equation}
where $\partial_i\,:=\,\frac{\partial}{\partial P^i}$. 

\noindent
 Without loss of generality, the following results are proven for the standard sequence
 $(\sigma_j)_{j=0}^{\infty}$ of infrared cutoffs. By simple arguments (see \cite{Pizzo}), limiting behavior as $\sigma\to0$  is shown to
 be ``sequence-independent".

\noindent
By analytic perturbation theory we have that
\begin{eqnarray}
& &\partial_i^2E^{\sigma_j}_{|\vP|}|_{\vP=P^{i}\hat{i}}\\
&=&1-2\langle \frac{1}{2\pi i}\oint_{\gamma_j}\frac{1}{H^{\sigma_j}_{\vP}-z_j}[P^{i}-(\beta^{\sigma_j})^{i}]\frac{1}{H^{\sigma_j}_{\vP}-z_j}dz_j\,\Psi_{\vP}^{\sigma_j},\,[P^{i}-(\beta^{\sigma_j})^{i}]\Psi_{\vP}^{\sigma_j}\rangle|_{\vP=P^{i}\hat{i}} \,, \nonumber
\end{eqnarray}
here $\Psi_{\vP}^{\sigma_j}$ is the normalized ground state eigenvector of $H_{\vP}^{\sigma_j}$.

\noindent
Next, we make use of the Bogoliubov transformation implemented by $W_{\sigma_j}(\vnabla
E_{\vP}^{\sigma_j})$ to show that
\begin{eqnarray}
& &\langle \frac{1}{2\pi i}\oint_{\gamma_j}\frac{1}{H^{\sigma_j}_{\vP}-z_j}[P^{i}-(\beta^{\sigma_j})^{i}]\frac{1}{H^{\sigma_j}_{\vP}-z_j}dz_j\,\Psi_{\vP}^{\sigma_j},\,[P^{i}-(\beta^{\sigma_j})^{i}]\Psi_{\vP}^{\sigma_j}\rangle \\
&=&\frac{1}{\|\Phi_{\vP}^{\sigma_j}\|^2}\langle \frac{1}{2\pi i}\oint_{\gamma_j}\frac{1}{K^{\sigma_j}_{\vP}-z_j}[P^{i}-W_{\sigma_j}(\vnabla
E_{\vP}^{\sigma_j})(\beta^{\sigma_j})^{i}W_{\sigma_j}^*(\vnabla
E_{\vP}^{\sigma_j})]\frac{1}{K^{\sigma_j}_{\vP}-z_j}dz_j\,\Phi_{\vP}^{\sigma_j}\,,\nonumber\\
& &\quad\quad\quad\quad\quad,\,[P^{i}-W_{\sigma_j}(\vnabla
E_{\vP}^{\sigma_j})(\beta^{\sigma_j})^{i}W_{\sigma_j}^*(\vnabla
E_{\vP}^{\sigma_j})]\Phi_{\vP}^{\sigma_j}\rangle \,,
\end{eqnarray}
where $\Phi_{\vP}^{\sigma_j}$ is the ground state eigenvector of
$K_{\vP}^{\sigma_j}$ (iteratively constructed in Section II).

\noindent
Recalling the definitions
\begin{equation}
\vec{\Pi}_{\vP}^{\sigma_j}\,:=\,W_{\sigma_j}(\vnabla
E_{\vP}^{\sigma_j})\vec{\beta}^{\sigma_j}W_{\sigma_j}^*(\vnabla
E_{\vP}^{\sigma_j})
-\langle W_{\sigma_j}(\vnabla
E_{\vP}^{\sigma_j})\vec{\beta}^{\sigma_j}W_{\sigma_j}^*(\vnabla
E_{\vP}^{\sigma_j}) \rangle_{\Omega_f}\,,
\end{equation}
\begin{equation}
\vec{\Gamma}^{\sigma_j}_{\vP}\,:=\,\vec{\Pi}_{\vP}^{\sigma_j}-\langle
  \vec{\Pi}_{\vP}^{\sigma_j} \rangle_{\Phi_{\vP}^{\sigma_j}}\,,
\end{equation}
and because of the identity (Feynman-Hellman, see (\ref{eq:III.36bis}))
\begin{eqnarray}
\langle \vec{\beta}^{\sigma_j} \rangle_{\psi_{\vP}^{\sigma_j}}&=&\vP-\vnabla
E_{\vP}^{\sigma_j}\\
&= &\langle
  \vec{\Pi}_{\vP}^{\sigma_j} \rangle_{\Phi_{\vP}^{\sigma_j}}+\langle W_{\sigma_j}(\vnabla
E_{\vP}^{\sigma_j})\vec{\beta}^{\sigma_j}W_{\sigma_j}^*(\vnabla
E_{\vP}^{\sigma_j}) \rangle_{\Omega_f}\,,
\end{eqnarray}
we find that
\begin{equation} 
P^{i}-W_{\sigma_j}(\vnabla
E_{\vP}^{\sigma_j})(\beta^{\sigma_j})^{i}W_{\sigma_j}^*(\vnabla
E_{\vP}^{\sigma_j})=-(\Gamma^{\sigma_j}_{\vP})^{i}+\partial_iE_{\vP}^{\sigma_j}\,;
\end{equation}
hence,
\begin{eqnarray}
& &\partial_i^2E^{\sigma_j}_{|\vP|}|_{\vP=P^{i}\hat{i}}\\
&=&1-2\frac{1}{\|\Phi_{\vP}^{\sigma_j}\|^2}\langle \frac{1}{2\pi i}\oint_{\gamma_j}\frac{1}{K^{\sigma_j}_{\vP}-z_j}[\partial_iE_{\vP}^{\sigma_j}-(\Gamma^{\sigma_j}_{\vP})^{i}]\frac{1}{K^{\sigma_j}_{\vP}-z_j}dz_j\,\Phi_{\vP}^{\sigma_j}\,,\nonumber\\
& &\quad\quad\quad\quad\quad\quad\quad\quad\quad\quad\,,\,[\partial_iE_{\vP}^{\sigma_j}-(\Gamma^{\sigma_j}_{\vP})^{i}]\Phi_{\vP}^{\sigma_j}\rangle|_{\vP=P^{i}\hat{i}} \,.
\end{eqnarray}

\noindent
Using the eigenvalue equation
$$K^{\sigma_j}_{\vP}\Phi_{\vP}^{\sigma_j}=E_{\vP}^{\sigma_j}\Phi_{\vP}^{\sigma_j},$$
the terms proportional to $(\partial_iE_{\vP}^{\sigma_j})^2$ and to the mixed terms -- i.e., proportional to the
product of $\partial_iE_{\vP}^{\sigma_j}$ and $(\Gamma^{\sigma_j}_{\vP})^{i}$
-- are seen to be identically $0$, because  the contour integral vanishes  for
each $i=1,2,3$; e.g., 
\begin{eqnarray}\label{eq:III.48}
\lefteqn{\oint_{\gamma_j}\langle \frac{1}{K^{\sigma_j}_{\vP}-z_j}[\partial_iE_{\vP}^{\sigma_j}]\frac{1}{K^{\sigma_j}_{\vP}-z_j}\,\Phi_{\vP}^{\sigma_j},\,[\partial_iE_{\vP}^{\sigma_j}]\Phi_{\vP}^{\sigma_j}\rangle d\bar{z}_j\quad\quad}\quad\\
&=&\oint_{\gamma_j}\langle\Phi_{\vP}^{\sigma_j}\,,\,\Phi_{\vP}^{\sigma_j}\rangle \big(\frac{\partial_iE_{\vP}^{\sigma_j}}{E^{\sigma_j}_{\vP}-\bar{z_j}}\big)^2d\bar{z}_j=0\,.\nonumber
\end{eqnarray}


\noindent
It follows that
\begin{eqnarray}
& &\partial_i^2E^{\sigma_j}_{|\vP|}|_{\vP=P^{i}\hat{i}}\\
&=&1+\frac{1}{\pi i}\oint_{\gamma_j}d\bar{z}_j\langle \frac{1}{K^{\sigma}_{\vP}-z_j}\,(\Gamma^{\sigma_j}_{\vP})^{i}\,\frac{1}{K^{\sigma}_{\vP}-z_j}\,\frac{\Phi_{\vP}^{\sigma_j}}{\|\Phi_{\vP}^{\sigma_j}\|},\,(\Gamma^{\sigma_j}_{\vP})^{i}\,\frac{\Phi_{\vP}^{\sigma_j}}{\|\Phi_{\vP}^{\sigma_j}\|}\rangle|_{\vP=P^{i}\hat{i}}\,\quad\\
&=&1+\frac{1}{\pi i}\oint_{\gamma_j}d\bar{z}_j\frac{1}{E^{\sigma_j}_{\vP}-\bar{z_j}}\langle \,(\Gamma^{\sigma_j}_{\vP})^{i}\,\frac{1}{K^{\sigma_j}_{\vP}-z_j}\,\,(\Gamma^{\sigma_j}_{\vP})^{i}\,\frac{\Phi_{\vP}^{\sigma_j}}{\|\Phi_{\vP}^{\sigma_j}\|}\,,\,\frac{\Phi_{\vP}^{\sigma_j}}{\|\Phi_{\vP}^{\sigma_j}\|}\rangle |_{\vP=P^{i}\hat{i}}\,.\quad\quad\quad\quad\label{eq:IV.48}
\end{eqnarray}
\\
We are now ready for the key estimate.

\begin{lemma}\label{LemmaIII.1}
The estimate below holds true ($j\in \NN$):
\begin{eqnarray}\label{eq:IV.19}
\lefteqn{\big|\oint_{\gamma_{j-1}}\big\langle (\Gamma_{\vP}^{\sigma_{j-1}})^{i}\,\frac{\Phi_{\vP}^{\sigma_{j-1}}}{\|\Phi_{\vP}^{\sigma_{j-1}}\|}\,,\,\frac{1}{K_{\vP}^{\sigma_{j-1}}-\bar{z}_{j-1}}(\Gamma_{\vP}^{\sigma_{j-1}})^{i}\,\frac{\Phi_{\vP}^{\sigma_{j-1}}}{\|\Phi_{\vP}^{\sigma_{j-1}}\|}\big \rangle \frac{1}{E^{\sigma_{j-1}}_{\vP}-\bar{z}_{j-1}} d\bar{z}_{j-1}}\,\nonumber\\
& &\quad-\oint_{\gamma_{j}}\big\langle
(\Gamma_{\vP}^{\sigma_j})^{i}\,\frac{\Phi_{\vP}^{\sigma_j}}{\|\Phi_{\vP}^{\sigma_j}\|}\,,\,\frac{1}{K_{\vP}^{\sigma_j}-\bar{z}_{j}}(\Gamma_{\vP}^{\sigma_j})^{i}\,\frac{\Phi_{\vP}^{\sigma_j}}{\|\Phi_{\vP}^{\sigma_j}\|}\big
\rangle \frac{1}{E^{\sigma_j}_{\vP}-\bar{z_j}} d\bar{z}_{j}\big|\leq \epsilon^{j(1-2\delta)}\,,\quad\quad\quad\label{eq:IV.49}
\end{eqnarray}
for any $0<\delta(<1/2)$, and for $\alpha$ and $\epsilon$ small enough depending on $\delta$.
\end{lemma}

\noindent
\emph{Proof.}

\noindent
By unitarity of $W_{\sigma_{j}}(\vnabla E_{\vP}^{\sigma_{j-1}})W_{\sigma_{j}}^*(\vnabla E_{\vP}^{\sigma_{j}})$, 
\begin{eqnarray}
\lefteqn{\oint_{\gamma_{j}}\big\langle
(\Gamma_{\vP}^{\sigma_j})^{i}\,\frac{\Phi_{\vP}^{\sigma_j}}{\|\Phi_{\vP}^{\sigma_j}\|}\,,\,\frac{1}{K_{\vP}^{\sigma_j}-\bar{z}_{j}}(\Gamma_{\vP}^{\sigma_j})^{i}\,\frac{\Phi_{\vP}^{\sigma_j}}{\|\Phi_{\vP}^{\sigma_j}\|}\big
\rangle \frac{1}{E^{\sigma_j}_{\vP}-\bar{z}_j} d\bar{z}_{j}}\\
&=&\oint_{\gamma_{j}}\big\langle (\hat{\Gamma}_{\vP}^{\sigma_j})^{i}\,\frac{\hat{\Phi}_{\vP}^{\sigma_{j}}}{\|\hat{\Phi}_{\vP}^{\sigma_{j}}\|}\,,\,\frac{1}{\hat{K}_{\vP}^{\sigma_j}-\bar{z}_{j}}(\hat{\Gamma}_{\vP}^{\sigma_j})^{i}\,\frac{\hat{\Phi}_{\vP}^{\sigma_{j}}}{\|\hat{\Phi}_{\vP}^{\sigma_{j}}\|}\big \rangle \frac{1}{E^{\sigma_{j}}_{\vP}-\bar{z}_{j}}d\bar{z}_{j}\,.
\end{eqnarray}
By assumption, $\alpha$ is so small that the Neumann series expansions of the
resolvents below converge in $\cF^{b}_{\sigma_j}$:
\begin{equation}
\frac{1}{\hat{K}_{\vP}^{\sigma_j}-\bar{z}_{j}}=\frac{1}{K_{\vP}^{\sigma_{j-1}}-\bar{z}_{j}}+\Sigma_{1}^{\infty}(K_{\vP}^{\sigma_{j-1}},\bar{z}_{j})\,,
\end{equation}
\begin{equation}
\frac{1}{E^{\sigma_{j}}_{\vP}-\bar{z}_{j}}=\frac{1}{E^{\sigma_{j-1}}_{\vP}-\bar{z_{j}}}+\Delta(E^{\sigma_{j-1}}_{\vP},\bar{z_{j}})\,,
\end{equation}
where:
\begin{eqnarray}
& &\Sigma_{1}^{\infty}(K_{\vP}^{\sigma_{j-1}},\bar{z}_{j})\\
&:=&\sum_{l=1}^{\infty}\frac{1}{K_{\vP}^{\sigma_{j-1}}-\bar{z}_{j}}[-(\Delta K_{\vP}|^{\sigma_{j-1}}_{\sigma_j}+\hat{\cE}_{\vP}^{\sigma_j}-\cE_{\vP}^{\sigma_{j-1}})\frac{1}{K_{\vP}^{\sigma_{j-1}}-\bar{z}_{j}}]^l\,,\nonumber
\end{eqnarray}
and $\Delta K_{\vP}|^{\sigma_{j-1}}_{\sigma_j}$ is defined in Eq. (\ref{eq:III.53});
\begin{equation}
\Delta(E^{\sigma_{j-1}}_{\vP},\bar{z}_{j}):=\frac{1}{E^{\sigma_{j}}_{\vP}-\bar{z}_{j}}(E^{\sigma_{j-1}}_{\vP}-E^{\sigma_{j}}_{\vP})\frac{1}{E^{\sigma_{j-1}}_{\vP}-\bar{z}_{j}}\,.
\end{equation}
We proceed by using the obvious identity:
\begin{eqnarray}
& &\oint_{\gamma_{j}}\big\langle (\hat{\Gamma}_{\vP}^{\sigma_j})^{i}\,\frac{\hat{\Phi}_{\vP}^{\sigma_{j}}}{\|\hat{\Phi}_{\vP}^{\sigma_{j}}\|}\,,\,\frac{1}{\hat{K}_{\vP}^{\sigma_j}-\bar{z}_{j}}(\hat{\Gamma}_{\vP}^{\sigma_j})^{i}\,\frac{\hat{\Phi}_{\vP}^{\sigma_{j}}}{\|\hat{\Phi}_{\vP}^{\sigma_{j}}\|}\big \rangle\frac{1}{E^{\sigma_{j}}_{\vP}- \bar{z}_{j}} d\bar{z}_{j}\\
&=&\oint_{\gamma_{j}}\big\langle (\hat{\Gamma}_{\vP}^{\sigma_j})^{i}\,\frac{\hat{\Phi}_{\vP}^{\sigma_{j}}}{\|\hat{\Phi}_{\vP}^{\sigma_{j}}\|}\,,\,\frac{1}{K_{\vP}^{\sigma_{j-1}}-\bar{z}_{j}}(\hat{\Gamma}_{\vP}^{\sigma_j})^{i}\,\frac{\hat{\Phi}_{\vP}^{\sigma_{j}}}{\|\hat{\Phi}_{\vP}^{\sigma_{j}}\|}\big \rangle\frac{1}{E^{\sigma_{j-1}}_{\vP}-\bar{z}_{j}} d\bar{z}_{j}\,\label{eq:IV.57}\\
& &+\oint_{\gamma_{j}}\big\langle (\hat{\Gamma}_{\vP}^{\sigma_j})^{i}\,\frac{\hat{\Phi}_{\vP}^{\sigma_{j}}}{\|\hat{\Phi}_{\vP}^{\sigma_{j}}\|}\,,\,\Sigma_{1}^{\infty}(K_{\vP}^{\sigma_{j-1}},\bar{z}_{j})(\hat{\Gamma}_{\vP}^{\sigma_j})^{i}\,\frac{\hat{\Phi}_{\vP}^{\sigma_{j}}}{\|\hat{\Phi}_{\vP}^{\sigma_{j}}\|}\big \rangle \frac{1}{E^{\sigma_{j-1}}_{\vP}-\bar{z}_{j}} d\bar{z}_{j}\,\,\,\,\quad\quad\label{eq:IV.58}\\
& & +\oint_{\gamma_{j}}\big\langle (\hat{\Gamma}_{\vP}^{\sigma_j})^{i}\,\frac{\hat{\Phi}_{\vP}^{\sigma_{j}}}{\|\hat{\Phi}_{\vP}^{\sigma_{j}}\|}\,,\,\frac{1}{\hat{K}_{\vP}^{\sigma_{j}}-\bar{z}_{j}}(\hat{\Gamma}_{\vP}^{\sigma_j})^{i}\,\frac{\hat{\Phi}_{\vP}^{\sigma_{j}}}{\|\hat{\Phi}_{\vP}^{\sigma_{j}}\|}\big \rangle\Delta(E^{\sigma_{j-1}}_{\vP},\bar{z}_{j})d\bar{z}_{j}\,.\label{eq:IV.59}
\end{eqnarray}
Each of the expressions (\ref{eq:IV.57}) and (\ref{eq:IV.58})  can be rewritten by adding and subtracting
$(\Gamma_{\vP}^{\sigma_{j-1}})^{i}\frac{\Phi_{\vP}^{\sigma_{j-1}}}{\|\Phi_{\vP}^{\sigma_{j-1}}\|}$. For
(\ref{eq:IV.57}) we get
\begin{eqnarray}
& &(\ref{eq:IV.57})\nonumber\\
& =& \oint_{\gamma_{j}}\big\langle (\Gamma_{\vP}^{\sigma_{j-1}})^{i}\frac{\Phi_{\vP}^{\sigma_{j-1}}}{\|\Phi_{\vP}^{\sigma_{j-1}}\|}\,,\,\frac{1}{K_{\vP}^{\sigma_{j-1}}-\bar{z}_{j}}(\Gamma_{\vP}^{\sigma_{j-1}})^{i}\frac{\Phi_{\vP}^{\sigma_{j-1}}}{\|\Phi_{\vP}^{\sigma_{j-1}}\|}\big \rangle \frac{1}{E^{\sigma_{j-1}}_{\vP}-\bar{z}_{j}}d\bar{z}_{j}\quad\quad\quad\quad\label{eq:IV.61}\\
& & +\oint_{\gamma_{j}}\big\langle
(\hat{\Gamma}_{\vP}^{\sigma_j})^{i}\frac{\hat{\Phi}_{\vP}^{\sigma_{j}}}{\|\hat{\Phi}_{\vP}^{\sigma_{j}} \|}-(\Gamma_{\vP}^{\sigma_{j-1}})^{i}\frac{\Phi_{\vP}^{\sigma_{j-1}}}{\|\Phi_{\vP}^{\sigma_{j-1}}\|}\,,\,\label{eq:IV.62}\\
& &\quad\quad\quad, \,\frac{1}{K_{\vP}^{\sigma_{j-1}}-\bar{z}_{j}}[(\hat{\Gamma}_{\vP}^{\sigma_j})^{i}\frac{\hat{\Phi}_{\vP}^{\sigma_{j}}}{\|\hat{\Phi}_{\vP}^{\sigma_{j}} \|}-(\Gamma_{\vP}^{\sigma_{j-1}})^{i}\frac{\Phi_{\vP}^{\sigma_{j-1}}}{\|\Phi_{\vP}^{\sigma_{j-1}}\|}]\big \rangle\frac{1}{E^{\sigma_{j-1}}_{\vP}-\bar{z}_{j}} d\bar{z}_{j}\nonumber\\
& & +\oint_{\gamma_{j}}\big\langle (\hat{\Gamma}_{\vP}^{\sigma_j})^{i}\frac{\hat{\Phi}_{\vP}^{\sigma_{j}}}{\|\hat{\Phi}_{\vP}^{\sigma_{j}} \|}-(\Gamma_{\vP}^{\sigma_{j-1}})^{i}\frac{\Phi_{\vP}^{\sigma_{j-1}}}{\|\Phi_{\vP}^{\sigma_{j-1}}\|}\,,\label{eq:IV.63}\\
& &\quad\quad\quad,\,\frac{1}{K_{\vP}^{\sigma_{j-1}}-\bar{z}_{j}}(\Gamma_{\vP}^{\sigma_{j-1}})^{i}\frac{\Phi_{\vP}^{\sigma_{j-1}}}{\|\Phi_{\vP}^{\sigma_{j-1}}\|}\big \rangle\frac{1}{E^{\sigma_{j-1}}_{\vP}-\bar{z}_{j}} d\bar{z}_{j}\,\nonumber\\
& & +\oint_{\gamma_{j}}\big\langle (\Gamma_{\vP}^{\sigma_{j-1}})^{i}\frac{\Phi_{\vP}^{\sigma_{j-1}}}{\|\Phi_{\vP}^{\sigma_{j-1}}\|}\,,\label{eq:IV.64}\\
& &\quad\quad\quad,\,\frac{1}{K_{\vP}^{\sigma_{j-1}}-\bar{z}_{j}}[(\hat{\Gamma}_{\vP}^{\sigma_j})^{i}\frac{\hat{\Phi}_{\vP}^{\sigma_{j}}}{\|\hat{\Phi}_{\vP}^{\sigma_{j}}\|}-(\Gamma_{\vP}^{\sigma_{j-1}})^{i}\frac{\Phi_{\vP}^{\sigma_{j-1}}}{\|\Phi_{\vP}^{\sigma_{j-1}}\|}]\big \rangle\frac{1}{E^{\sigma_{j-1}}_{\vP}-\bar{z}_{j}} d\bar{z}_{j}\,\nonumber\,.
\end{eqnarray}
The difference in Eq. (\ref{eq:IV.49}) corresponds to the sum of the terms
(\ref{eq:IV.58})-(\ref{eq:IV.59}) and of the terms
(\ref{eq:IV.62})-(\ref{eq:IV.64}). In fact, (\ref{eq:IV.61}) corresponds to the first term in (\ref{eq:IV.49}) after  a contour deformation from $\gamma_{j-1}$ to $\gamma_j$. 

\noindent
The sum of the remainder terms (\ref{eq:IV.58}), (\ref{eq:IV.59}), and
(\ref{eq:IV.62})-(\ref{eq:IV.64}) can be
bounded by $\epsilon^{j(1-2\delta)}$,  for $\cR_0$ and $\alpha$ small enough  but independent
of $j$, for any $\vP\in\cS$. (We recall that $\cR_0$ can be taken arbitrarily
small,  provided $\alpha$ is small enough). The details are as follows.
\begin{itemize}
\item
For (\ref{eq:IV.62})-(\ref{eq:IV.64})  use the following inequalities
\begin{eqnarray}
&
&\big\|\big(\frac{1}{K_{\vP}^{\sigma_{j-1}}-\bar{z}_{j}}\big)(\Gamma_{\vP}^{\sigma_{j-1}})^{i}\,\Phi_{\vP}^{\sigma_{j-1}}\big\| \leq \cO(\frac{\cR_0^{\frac{1}{2}}}{\alpha^{\frac{1}{4}}\,
  \epsilon^{(j-1)\delta}})\,,\label{eq:IV.65}\\
&
&\big\|[(\hat{\Gamma}_{\vP}^{\sigma_{j}})^{i}-(\Gamma_{\vP}^{\sigma_{j-1}})^{i}\big]\,\hat{\Phi}_{\vP}^{\sigma_{j}}\big\| 
\leq \cO(\alpha^{\frac{1}{4}} \epsilon^{j(1-\delta)})\,,\label{eq:IV.66}\\
&
&\big\|\frac{1}{K_{\vP}^{\sigma_{j-1}}-\bar{z}_{j}} \big\|_{\cF_{\sigma_j}} 
\leq \cO(\frac{1}{\epsilon^{j}})\,,\\
&
&\big\|\big(\frac{1}{K_{\vP}^{\sigma_{j-1}}-\bar{z}_{j}}\big)(\Gamma_{\vP}^{\sigma_{j-1}})^{i}\,(\hat{\Phi}_{\vP}^{\sigma_{j}}-\Phi_{\vP}^{\sigma_{j-1}})\big\| \leq \cO(\frac{\alpha^{\frac{1}{4}} \epsilon^{j(1-\delta)}}{\epsilon^{j}})\,,\\
&
&\big\|\hat{\Phi}_{\vP}^{\sigma_{j}}-\Phi_{\vP}^{\sigma_{j-1}}\big\| \leq \alpha^{\frac{1}{4}} \epsilon^{j(1-\delta)}\,.\label{eq:IV.68}
\end{eqnarray}
In order to derive the inequality in Eq. (\ref{eq:IV.66}), one uses Eqs. (\ref{eq:III.63}), (\ref{eq:III.73}), and (\ref{eq:III.57})-(II.58).
\item
For (\ref{eq:IV.58}), after adding and subtracting
$(\Gamma_{\vP}^{\sigma_{j-1}})^{i}\frac{\Phi_{\vP}^{\sigma_{j-1}}}{\|\Phi_{\vP}^{\sigma_{j-1}}\|}$,
one also has to use that
\begin{equation}
\|[-(\Delta
K_{\vP}|^{\sigma_{j-1}}_{\sigma_j}+\hat{\cE}_{\vP}^{\sigma_j}-\cE_{\vP}^{\sigma_{j-1}})]\frac{1}{K_{\vP}^{\sigma_{j-1}}-z_{j}}(\Gamma_{\vP}^{\sigma_{j-1}})^{i}\Phi_{\vP}^{\sigma_{j-1}}\|\leq
\cO(\alpha^{\frac{1}{2}}\epsilon^{j-1}\,\frac{\cR_0^{\frac{1}{4}}}{\alpha^{\frac{1}{4}}\,
  \epsilon^{(j-1)\delta}})\,;
\end{equation}
\item
To bound (\ref{eq:IV.59}), note that
\begin{eqnarray}
(\ref{eq:IV.59})&=&-2\pi i\big\langle (\hat{\Gamma}_{\vP}^{\sigma_j})^{i}\,\frac{\hat{\Phi}_{\vP}^{\sigma_{j}}}{\|\hat{\Phi}_{\vP}^{\sigma_{j}}\|}\,,\,\frac{1}{\hat{K}_{\vP}^{\sigma_{j}}-E_{\vP}^{\sigma_j}}(\hat{\Gamma}_{\vP}^{\sigma_j})^{i}\,\frac{\hat{\Phi}_{\vP}^{\sigma_{j}}}{\|\hat{\Phi}_{\vP}^{\sigma_{j}}\|}\big \rangle\\
& &+2\pi i\big\langle
(\hat{\Gamma}_{\vP}^{\sigma_j})^{i}\,\frac{\hat{\Phi}_{\vP}^{\sigma_{j}}}{\|\hat{\Phi}_{\vP}^{\sigma_{j}}\|}\,,\,\frac{1}{\hat{K}_{\vP}^{\sigma_{j}}-E_{\vP}^{\sigma_{j-1}}}(\hat{\Gamma}_{\vP}^{\sigma_j})^{i}\,\frac{\hat{\Phi}_{\vP}^{\sigma_{j}}}{\|\hat{\Phi}_{\vP}^{\sigma_{j}}\|}\big
\rangle\,\quad\quad\quad\\
&= &  2\pi i\big\langle (\hat{\Gamma}_{\vP}^{\sigma_j})^{i}\,\frac{\hat{\Phi}_{\vP}^{\sigma_{j}}}{\|\hat{\Phi}_{\vP}^{\sigma_{j}}\|}\,,\,\frac{(E_{\vP}^{\sigma_{j-1}}-E_{\vP}^{\sigma_j})}{\hat{K}_{\vP}^{\sigma_{j}}-E_{\vP}^{\sigma_j}}\,\frac{1}{\hat{K}_{\vP}^{\sigma_{j}}-E_{\vP}^{\sigma_{j-1}}}(\hat{\Gamma}_{\vP}^{\sigma_j})^{i}\,\frac{\hat{\Phi}_{\vP}^{\sigma_{j}}}{\|\hat{\Phi}_{\vP}^{\sigma_{j}}\|}\big \rangle\quad\quad\quad\quad\quad\quad
\end{eqnarray}
where $|E_{\vP}^{\sigma_{j-1}}-E_{\vP}^{\sigma_{j}}|\leq
\cO(\alpha\,\epsilon^{j-1})$. Then use that $\epsilon\geq C\alpha^{\frac{1}{2}}$ and  the following
 inequality analogous to (\ref{eq:IV.65})
\begin{equation}
\big\|\big(\frac{1}{\hat{K}_{\vP}^{\sigma_{j}}-E_{\vP}^{\sigma_j}}\big)(\hat{\Gamma}_{\vP}^{\sigma_{j}})^{i}\,\hat{\Phi}_{\vP}^{\sigma_{j}}\big\| \leq \cO(\frac{\cR_0^{\frac{1}{2}}}{\alpha^{\frac{1}{4}}\,
  \epsilon^{j\delta}})\,.
\end{equation}
\end{itemize}
\QED

\begin{theorem}
For $\alpha$ small enough, $\frac{\partial^2E^{\sigma}_{|\vP|}}{(\partial|\vP|)^2}$ converges,  as $\sigma\to0$. The limiting function, $ \Sigma_{|\vP|}:=\lim_{\sigma\to0}\,\frac{\partial^2E^{\sigma}_{|\vP|}}{(\partial|\vP|)^2}$, is H\"older-continuous in $\vP\in\cS$ (for an exponent $\eta>0$). The limit
\begin{equation} \label{eq:IV.71}
\lim_{\alpha \to0}\,\Sigma_{|\vP|}=1\,
\end{equation}
holds true uniformly in $\vP\in\cS$.
\end{theorem}

\noindent
\emph{Proof}

\noindent
It is enough to prove the result for a fixed choice of a sequence $\{\sigma_j\}_{j=0}^{\infty}$. The estimate
in Lemma \ref{LemmaIII.1} implies the existence of $\lim_{j\to\infty}\,\partial_i^2E^{\sigma_j}_{|\vP|}|_{\vP=P^i\hat{i}}$.

\noindent
We now observe that $\partial_i^2E^{\sigma_0}_{|\vP|}|_{\vP=P^{i}\hat{i}}=1$ (see Eq. (\ref{eq:IV.48})),
because
\begin{equation}\label{eq:IV.72}
(\Gamma_{\vP}^{\sigma_{0}})^{i}\equiv (\vP^f)^{i}\quad,\quad\Phi_{\vP}^{\sigma_{0}}\equiv\Omega_f\,.
\end{equation}
According to the constraint in Eq. (\ref {eq:II.83b}),  we can take $\epsilon=\cO(\alpha^{\frac{1}{2}(1-\delta)})$ so that,  for $\alpha$ small enough, Lemma \ref{LemmaIII.1}  and  (\ref{eq:IV.72})  yield
\begin{equation}\label{eq:IV.73}
\big|\frac{1}{\pi i}\oint_{\gamma_j}d\bar{z}_j\frac{1}{E^{\sigma_j}_{\vP}-\bar{z}_j}  \langle \,\frac{\Phi_{\vP}^{\sigma_j}}{\|\Phi_{\vP}^{\sigma_j}\|}\,,\,(\Gamma^{\sigma_j}_{\vP})^{i}\,\frac{1}{K^{\sigma_j}_{\vP}-\bar{z}_j}\,\,(\Gamma^{\sigma_j}_{\vP})^{i}\,\frac{\Phi_{\vP}^{\sigma_j}}{\|\Phi_{\vP}^{\sigma_j}\|}\rangle|_{\vP=P^{i}\hat{i}}\big|<\cO(\alpha^{\frac{1}{2}(1-\delta)(1-2\delta)})\,,
\end{equation}
uniformly in $j\in\NN$.
Hence the limit (\ref{eq:IV.71}) follows. 

\noindent
The H\"older-continuity in $\vP$ of $\Sigma_{|\vP|}$ is a trivial consequence of the analyticity in $\vP\in\cS$ of $E^{\sigma}_{\vP}$,  for any $\sigma>0$, and of Lemma \ref{LemmaIII.1}; see \cite{Pizzo} for similar results.
\QED

\begin{corollary}
For $\alpha$ small enough, the function 
$E_{\vP}:=\lim_{\sigma\to0}E_{\vP}^{\sigma}$, $\vP\in\cS$,  is twice
differentiable, and 
\begin{equation}
\frac{\partial^2E_{|\vP|}}{(\partial |\vP|)^2}=\Sigma_{|\vP|}\,.
\end{equation}
\end{corollary}

\noindent
\emph{Proof}

\noindent
The result follows from the H\"older-continuity of $\Sigma_{|\vP|}$, of
$\lim_{\sigma\to0} \frac{\partial E_{|\vP|}^{\sigma}}{\partial|\vP|}$, and from the fundamental theorem of calculus applied
to the functions $E_{\vP}$ and $\lim_{\sigma\to0} \frac{\partial
  E_{|\vP|}^{\sigma}}{\partial|\vP|}$, because 
\begin{itemize}
\item
\begin{equation}
\frac{\partial E_{|\vP|}^{\sigma}}{\partial|\vP|} \quad \text{and}\quad
\frac{\partial^2 E_{|\vP|}^{\sigma}}{(\partial|\vP|)^2}
\end{equation}
converge pointwise, for $\vP\in \cS$, as $\sigma\to0$,
\item
\begin{equation}
\big|\frac{\partial
E_{|\vP|}^{\sigma}}{\partial|\vP|}\big|\quad \text{and}\quad\big|\frac{\partial^2
E_{|\vP|}^{\sigma}}{(\partial|\vP|)^2}\big|
\end{equation} 
are  uniformly bounded in $\sigma$, 
for all $\vP\in\cS$.
\end{itemize}
\QED

\subsubsection*{Acknowledgements}

The authors thank Thomas Chen for very useful discussions.


\begin{thebibliography}{10}





\bibitem{BachFroehlichPizzo}
V.~ Bach, J. ~Froehlich, A.~ Pizzo. 
\newblock  Infrared-finite algorithms in QED II. The expansion of the groundstate of an atom interacting with the quantized radiation field. 
\newblock {\em Adv. Math.  Volume 220, Issue 4, Pages 1023-1074}


\bibitem{ChenBachFroehlichSigal}
V.~Bach,T.~Chen, J.~Fr{\"{o}}hlich, and I.~M. Sigal.
\newblock The Renormalized Electron Mass in Non-Relativistic Quantum Electrodynamics.
\newblock http://arxiv.org/abs/math-ph/0507043


\bibitem{Chen}
T.~Chen.
\newblock  Infrared Renormalization in Nonrelativistic QED and Scaling Criticality. 
\newblock http://arxiv.org/abs/math-ph/0601010



\bibitem{ChFr}
T.~Chen, J.~Fr{\"{o}}hlich.
\newblock Coherent Infrared Representations in Nonrelativistic QED,
{\em To appear in Proc. Symp. Pure Math. (B. Simon 60-th Birthday Volume)}.  
\newblock http://arxiv.org/abs/math-ph/0601009 

\bibitem{ChenFroehlichPizzo1}
T.~Chen, J.~Fr{\"{o}}hlich, and A.~Pizzo.
\newblock Infraparticle Scattering States in Non-Relativistic QED: I. The
Bloch-Nordsieck paradigm
\newblock http://arxiv.org/abs/math-ph/07092493

\bibitem{ChenFroehlichPizzo2}
T.~Chen, J.~Fr{\"{o}}hlich, and A.~Pizzo.
\newblock Infraparticle Scattering States in Non-Relativistic QED: II. Mass
Shell Properties.
\newblock  {\em J. Math. Phys. 50, 012103 (2009); DOI:10.1063/1.3000088}

\bibitem{Fierz-Pauli}
M.~Fierz, W.~Pauli.
\newblock{\em Nuovo Cimento, 15 167 (1938)}







\bibitem{Hiroshima-Spohn}
F. Hiroshima, H. Spohn.
\newblock { Mass renormalization in non-relativistic quantum electrodynamics.}
\newblock {\em J. Math. Phys.. 46 (2005) (4)}

\bibitem{Hainzl-Seiringer}
C. Hainzl, R. Seiringer.
\newblock {Mass renormalization and energy level shift in non-relativistic QED.}
\newblock {\em Adv. Theor. Math. Phys. (6) (2003) (5), 847-871.}

\bibitem{Lieb-Loss}
E.T. Lieb, M. Loss.
\newblock{Self energy of electrons in non-perturbative QED.}
\newblock{\em Conference Moshe Flato 1999, vol I, Dijon,
  Math. Phys. Stud. vol. 21, Kluwer Acad. Publ., Dordrecht (2000),
  pp. 327-344.}

\bibitem{Lieb-Loss2}
E.T. Lieb, M. Loss.
\newblock{A bound on binding energies and mass renormalization in model of
  quantum electrodynamics.}
\newblock {\em J. Statist. Phys. 108 (2002)}


\bibitem{Pizzo}
A.~Pizzo.
\newblock One-particle (improper) states in {N}elson's massless model.
\newblock {\em Ann.~H.~Poincar{\'e}}, {\bf 4} (3), 439--486, 2003.

\bibitem{Pizzo2005}
A.~Pizzo.
\newblock Scattering of an \emph{Infraparticle}: The One-particle (improper) Sector in Nelson's massless model.
\newblock {\em Ann.~H.~Poincar{\'e}}, {\bf 4} (3), 439--486, 2003.

\bibitem{Schweber}
S.S.~Schweber.
\newblock An introduction to Relativistic Quantum Field Theory.
\newblock {\em Harper and Row, New York, 1961}.

\bibitem{Simon}
Reed Simon.
\newblock Methods of modern mathematical physics. Vol. I-II-III-IV
\newblock {\em Academic Press}.
\end{thebibliography}
\end{document}